\documentclass[10pt,journal,twocolumn]{IEEEtran}

\RequirePackage[hyphens]{url}

\usepackage{%
    amsmath,
    amsfonts,
    graphicx,
    algorithm,
    algorithmic,
    enumitem,
    booktabs,
    amsthm,
    colortbl,
    multirow,
    mathtools,
    float,
    array,
    makecell,
}

\usepackage{hyperref}
\hypersetup{
colorlinks=true,
linkcolor=blue,
citecolor=blue,
urlcolor=blue,
}

\DeclarePairedDelimiter\floor{\lfloor}{\rfloor}
\DeclareMathOperator*{\argmin}{argmin}

\definecolor{lightgray}{gray}{0.95}

\algsetup{linenosize=\normalsize,linenodelimiter=.}

\newtheorem{prop}{Proposition}

\begin{document}

\title{Subspace-based Representation and Learning for Phonotactic Spoken Language Recognition}

\author{
Hung-Shin~Lee,~\IEEEmembership{Student~Member,~IEEE},
Yu~Tsao,~\IEEEmembership{Member,~IEEE},
Shyh-Kang~Jeng,~\IEEEmembership{Senior~Member,~IEEE},
and~Hsin-Min~Wang,~\IEEEmembership{Senior~Member,~IEEE}
\thanks{
H.-S Lee and S.-K Jeng are with Department of Electrical Engineering, National Taiwan University, Taiwan, e-mail: hungshinlee@gmail.com.}
\thanks{
Y. Tsao is with Research Center for Information Technology Innovation, Academia Sinica, Taiwan, e-mail: yu.tsao@citi.sinica.edu.tw.}
\thanks{
H.-M Wang is with Institute of Information Science, Academia Sinica, Taiwan, e-mail: whm@iis.sinica.edu.tw.}}

\maketitle

\begin{abstract}
Phonotactic constraints can be employed to distinguish languages by representing a speech utterance as a multinomial distribution or phone events. In the present study, we propose a new learning mechanism based on subspace-based representation, which can extract concealed phonotactic structures from utterances, for language verification and dialect/accent identification. The framework mainly involves two successive parts. The first part involves subspace construction. Specifically, it decodes each utterance into a sequence of vectors filled with phone-posteriors and transforms the vector sequence into a linear orthogonal subspace based on low-rank matrix factorization or dynamic linear modeling. The second part involves subspace learning based on kernel machines, such as support vector machines and the newly developed subspace-based neural networks (SNNs). The input layer of SNNs is specifically designed for the sample represented by subspaces. The topology ensures that the same output can be derived from identical subspaces by modifying the conventional feed-forward pass to fit the mathematical definition of subspace similarity. Evaluated on the ``General LR'' test of NIST LRE 2007, the proposed method achieved up to 52\%, 46\%, 56\%, and 27\% relative reductions in equal error rates over the sequence-based PPR-LM, PPR-VSM, and PPR-IVEC methods and the lattice-based PPR-LM method, respectively. Furthermore, on the dialect/accent identification task of NIST LRE 2009, the SNN-based system performed better than the aforementioned four baseline methods.
\end{abstract}


\IEEEpeerreviewmaketitle

\section{Introduction}

\IEEEPARstart{S}{poken} language recognition (SLR) is a branch of audio classification where sound patterns extracted from raw waveform data are typically expressed by discrete symbol sequences or continuous-valued feature vectors. The performance of SLR tasks depends significantly on data representation \cite{Li2013}, which can be constructed from three levels of information: acoustic, phonetic, and phonotactic \cite{Ambikairajah2011,Zissman2001}. At the acoustic level, short-time frame-based features, such as mel-frequency cepstral coefficients (MFCCs) \cite{Davis1980}, linear prediction cepstral coefficients (LPCCs) \cite{Makhoul1975}, or shifted delta cepstral (SDC) features \cite{Torres-Carrasquillo2002}, are derived from speech signals. Probabilistic models, such as Gaussian mixture models and total variability models \cite{Wong2002,Martinez2011}, are generally used to form a vectorial representation for each utterance to be used by backend classifiers, such as support vector machines (SVMs) \cite{Campbell2006}. With the great success of deep learning in recent years, acoustic frames are directly fed into deep neural networks (DNNs) \cite{Lopez-Moreno2014,Richardson2015,Hautamaki2015,McCree2016,Ng2017,Srivastava2017}, recurrent neural networks (RNNs) \cite{Jiao2016}, and long short-term memory \cite{Gonzalez-Dominguez2014}.

At the phonetic level, phone log-likelihood ratio (PLLR) features are derived from the frame-by-frame phone posteriors provided by a neural network trained for frame-wise phone classification \cite{Diez2012}. One of the strengths of the PLLR is that it can directly combine phonetic and acoustic features \cite{DHaro2014}. In addition, bottleneck features (BNFs), i.e., feature streams generated from the linear bottleneck layer in a deep neural architecture, aim to bridge the gap between acoustic and phonetic levels of information \cite{Jiang2014,Matejka2014,McLaren2016,Yap2016}. Evolved BNFs were recently proposed by reformulating the goal to multilabel detection of articulatory attributes. The problem was solved using convolutional RNNs with a maximal figure-of-merit mathematical framework \cite{Cakir2017,Kukanov2020}.

By contrast, phonotactic approaches exploit a single or multiple phone recognizers to convert a speech utterance into one or several phone sequences to capture longer-term and higher-level phonetic information across languages. For example, the word-initial phone /s/ is more typically followed by non-nasal consonants in English than in German. In addition, the issue of rhoticity, whether /r/ is sounded when it follows a vowel, is a phonotactic constraint to identify whether the speaker is from England or the West Country in the United Kingdom \cite{Upton2012}. Therefore, if the phonotactic constraint can be sufficiently described for each utterance, the characteristics of each language will be well modeled and better recognition performance will be obtained. In this study, we intend to focus on the use of phonotactic (phonetic-contextual) information, which has been proven to be one of the most effective and convincible clues for SLR \cite{Li2013,Zissman1996}.

\textbf{Distributional and vectorial representations.} Various implementations have been proposed for handling phonotactic information. In these implementations, a phone recognizer is used to tokenize an utterance into a phone sequence. Some researchers have employed phone \textit{n}-gram modeling to capture the patterns of phone sequences, where each target language or each utterance can be described as a multinomial \textit{distribution} by using their phone \textit{n}-gram statistics. Thus, the similarity between a target language and an utterance can be measured by their negative cross entropy (or KL divergence) \cite{Yan1995} or based on the relative positions of \textit{n}-grams \cite{DeCordoba2007}, where discriminative \textit{n}-grams are selected to reduce the search space. Rather than using only the best phone sequence, Gauvain \textit{et al.} used the posterior probabilities provided by the phone lattice as soft counts for \textit{n}-gram modeling and obtained good results \cite{Gauvain2004}.

Stemming from the vector space modeling (VSM) framework in the field of information retrieval, Li \textit{et al.} arranged the phone \textit{n}-gram statistics of training and test samples in terms of \textit{n}-gram counts or term frequency-inverse document frequency (TF-IDF) weights into high-dimensional vectors \cite{Li2007}. They built a composite feature \textit{vector} for each utterance by concatenating the vectorized statistics from multiple recognizers and applied the vectors to backend classifiers. Based on the VSM framework, the research focus on phonotactic SLR was shifted to address two main issues: to obtain more discriminative features in the VSM-based representation for classification and to extract a more compact yet informative representation. For the first issue, Richardson and Campbell used SVMs to select key \textit{n}-gram terms \cite{Richardson2008}. Tong \textit{et al.} expanded each existing phone decoder to multiple target-oriented phone recognizers by selecting a subset of phones for each target language that can best discriminate the target language from other languages \cite{Tong2009}. Penagarikano \textit{et al.} used time alignment information by considering time-synchronous cross-decoder phone co-occurrences and defined a new concept of multiphone labels to integrate the contributions afforded by multiple decoders in a frame-by-frame manner \cite{Penagarikano2011}. For the second issue, some researchers have attempted to apply unsupervised dimension reduction approaches, such as latent semantics analysis (LSA) \cite{Li2007} and principal component analysis (PCA) \cite{Mikolov2010} on the original VSM-based feature vectors to render the classification task more efficient and to avoid the curse of dimensionality when training data are deficient. In addition, owing to the superior performance in the speaker recognition field \cite{Dehak2011}, the idea of i-vectors has been introduced into phonotactic SLR, where each high-dimensional feature vector is projected onto a low-dimensional subspace, forming a compact vector. For instance, inspired by the probabilistic model proposed in \cite{Maas2010}, Soufifar \textit{et al.} used the subspace multinomial model (SMM) along with the maximum likelihood criterion to effectively represent the information contained in \textit{n}-grams \cite{Soufifar2011}. Its latest variant that introduces sparsity constraints is available in \cite{Kesiraju2016}.

\textbf{Subspace-based representations.} In addition to the aforementioned representations, in our previous work, we assumed that the phonotactic information in an utterance can be distilled maximally using phone decoders and borne by a \textit{subspace}, i.e., a collection of mutually orthogonal vectors lying on a specific manifold \cite{Shih2012,Lee2013}. This type of data representation was originally proposed by Oja and Kohonen for characterizing a class composed of multiple samples \cite{Oja1988}. It has been developed into a broader field that considers the mutuality of subspaces in a kernel or nonkernel manner \cite{Yamaguchi1998,Tsuda1999}. Subsequently, it was successfully applied to action recognition \cite{Harandi2013}, image search \cite{Wang2011}, and visual sequence learning \cite{Turaga2011}. However, to the best of our knowledge, it has not been applied to speech processing. The subspace-based framework for SLR may be able to more fully utilize phonetic information provided by phone decoders. This is because all phones, not limited to those in the single most likely decoded sequence, can be considered and used to compensate for information loss due to inaccurate phone decoders. This idea is similar to using phone lattices \cite{Gauvain2004} or posteriogram-based \textit{n}-gram counts \cite{DHaro2012}. Furthermore, the order of \textit{n}-grams or the length of phonotactic constraints are no longer limited by the exponentially increasing memory size required to model phone sequences. Therefore, the contextual relationship among adjacent phones can be sustained with a linear growth rate of space complexity in data representation via vectorial concatenation and matrix factorization \cite{Shih2012} or dynamic linear modeling \cite{Lee2013} for subspace formulation.

For two subspaces residing in a common Euclidean space, many metrics can be used to evaluate their similarity or distance based on their principal angles (or canonical correlations) \cite{Bjorck1973,DeCock2002,Zuccon2009}. As described in \cite{Risteski2001}, the notion of principal angles might not be the best geometrical definition of \textit{angles} between two subspaces. However, it can still contribute to useful distance functions, some of which are suitable for kernel machines, such as kernel linear discriminant analysis \cite{Scholkopf2001} and SVMs \cite{Hamm2008}, for classification. At this point, all subspaces must be on the same Grassmann manifold, which is a set of fixed-dimensional subspaces in the Euclidean space. Therefore, the kernel functions used in kernel machines or neural networks must be designed in a Grassmannian manner and characterized by two subspaces.

\textbf{Contributions.} In this study, we attempt to extend and rectify our prior works on the NIST LRE corpora \cite{Shih2012,Lee2013} and propose a more integrated subspace-based framework for SLR by threading three key components.

\begin{enumerate}[label=\arabic*),leftmargin=*]
\item
\textbf{Phonetic feature extraction} derives as much phonetic information as possible by multiple language-dependent phone recognizers for better robustness and generalization.
\item
\textbf{Subspace construction} transforms an utterance into subspace-based representations by considering its phonotactic information. In addition to orthogonality of a subspace, which is a necessary condition that can be easily achieved through singular value decomposition (SVD), we introduce a supplementary constraint of sparsity for \textit{orthogonal dictionary learning}.
\item
\textbf{Subspace learning} involves two types of classifiers, namely SVMs and a newly developed model called the subspace-based neural networks (SNNs). The final score fusion is no longer needed because the subspaces of an utterance generated by parallel phone recognizers are used as multiple inputs by the SNNs. Any two nearly identical input subspaces will get similar outputs. This means that the mapping between the input layer and output layer of the SNNs is mathematically functional, so it is feasible to optimize the entire model using back propagation \cite{Rumelhart1986}.
\end{enumerate}

\textbf{Organization.} The remainder of this paper is organized as follows. In Section \ref{sec:phonetic_feature_extraction}, we demonstrate the method to leverage phone decoders for gathering phonotactic information in each utterance. Section \ref{sec:subspace_construction} presents several methods for subspace construction. Section \ref{sec:subspace_learning} describes the learning mechanisms for subspaces. We describe our experiments and report the results on three NIST LRE corpora in Section \ref{sec:experiments}. Section \ref{sec:conclusions} concludes the paper.

\begin{figure}[!t]
\centering
\includegraphics[width=0.48\textwidth]{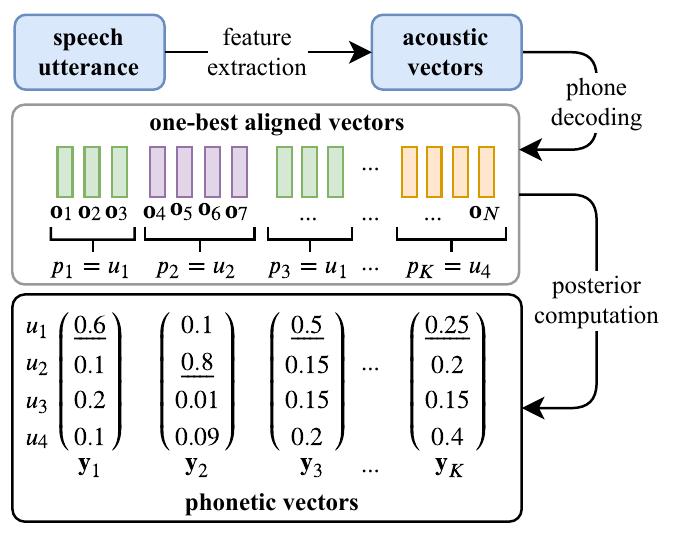}
\vspace{-10pt}
\caption{Phonetic vector extraction for an SLR system with a phonetic repertoire containing only four phones: $u_1$, $u_2$, $u_3$, and $u_4$.}
\label{fig:pv}
\vspace{-15pt}
\end{figure}

\section{Phonetic Feature Extraction}
\label{sec:phonetic_feature_extraction}
For the sequence of acoustic vectors, $\mathbf{o}_1,\dotsc,\mathbf{o}_T$, of an utterance, a phone decoder yields the best phone sequence $p_1,\dotsc,p_K$ and phone/frame alignment information. The basic idea of our phonotactic data representation is to use the best phone sequence provided by the phone decoder as a type of clustering in the acoustic vectors in a phonetic manner. As shown by the example in Fig. \ref{fig:pv}, after phone decoding, acoustic vectors $\mathbf{o}_1,\dotsc,\mathbf{o}_3$ are aligned to phone $u_1$, whereas $\mathbf{o}_4,\dotsc, \mathbf{o}_7$ belong to $u_2$. The acoustic vectors in a phone segment are further used to derive a more meaningful phonetic vector comprising the posterior probabilities of individual phone models. Consequently, each phone segment $p_k$ is represented by a phonetic vector $\mathbf{y}_k$, whose dimension is the size of the phone set.

The concept of phonetic vectors is that each utterance is compressed in length from $T$ acoustic frames to $K$ phonetic vectors ($K<T$). Each decoder need not be a highly accurate phone recognizer, but a \textit{sound} tokenizer. Moreover, the \textit{sound} boundary need not be consistent with the ground truth if it exists, even though the decoded \textit{sound} pattern might be inexplicable. The task of utilizing uncertainty information and learning patterns of phonotactic constraints is consigned to the procedure of contextualizing phonetic vectors, which is one of the key components in constructing a subspace.

\section{Subspace Construction}
\label{sec:subspace_construction}
A traditional subspace-based learning method is based on separately extracting the most characteristic properties of each class. Specifically, a subspace is represented or spanned by a set of vectors constructed from the feature vectors of each class \cite{Oja1988}. The learning mechanism depends on the similarity (or distance) measure between a class subspace and a sample vector. However, most common subspace-relevant methods for SLR, such as the i-vector and PCA, do not belong to this definition. On the contrary, their goal is to derive a coordinate representation (or a vectorial point) for each utterance in a single or joint lower-dimensional linear space. In our proposed framework, each utterance is a subspace. Using techniques such as \textit{phonetic vector contextualization} and \textit{orthogonal pursuit}, each subspace can be constructed by considering the salient phonotactic structure and the orthogonality, which is necessary for measuring the similarity between two subspaces.

\begin{figure}[!t]
\centering
\includegraphics[width=0.48\textwidth]{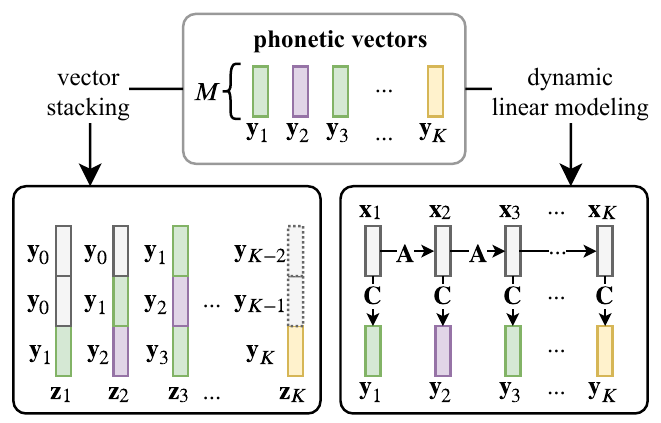}
\vspace{-10pt}
\caption{(Left) Vector stacking for a sequence of $K$ phonetic vectors with size $M$, where the context order is set to 3, and padding with the first vector is adopted. (Right) Dynamic linear modeling for deriving the temporal structure of the phonetic vectors, in terms of linear operators $\mathbf{A}$ and $\mathbf{C}$.}
\label{fig:pv_context}
\vspace{-15pt}
\end{figure}

\subsection{Phonetic Vector Contextualization}
\label{sec:phonetic_vector_contextualization}
We propose two methods to characterize the contextual relationship between adjacent phonetic vectors in an utterance.

\subsubsection{Vector Stacking}
\label{sec:vector_stacking}
Different methods can be used to acquire acoustic dynamics from filter-bank outputs by concatenating adjacent vectors into a larger one that contains longer-term information \cite{Beulen1995,Lee2008}. Analogously, we vertically stacked each phonetic vector up with its proximal neighbors, forming a higher-dimensional vector \cite{Shih2012}. A similar approach was adopted in \cite{Han2012}. As shown in Fig. \ref{fig:pv_context} for example, the context order $n$ possesses the similar idea of the notation $n$ in ``\textit{n}-gram,'' i.e., the larger the value of $n$, the longer can the phonotactic pattern be described. If we set $n$ to 3, the phonetic vector $\mathbf{y}_k$ is stacked with its two preceding vectors $\mathbf{y}_{k-2}$ and $\mathbf{y}_{k-1}$, forming a context-dependent super-vector $\mathbf{z}_k$. Regarding the phonetic vectors near the beginning of the utterance, such as $\mathbf{y}_1$ and $\mathbf{y}_2$, the non-existing historical vectors can be filled up by zero-padding (used in our experiments) or a duplication of the first vector. Therefore, for a phone set of size $M$, the total number of phonotactic features (i.e., number of \textit{tri}gram combinations) to represent an utterance with $K$ phones will reach $M^3$ under the \textit{n}-gram-based framework. However, in vector stacking, it will only be $3\times M\times K$, which is much smaller than $M^3$, as most speech utterances in practice are less than 30 s.

\subsubsection{Dynamic Linear Modeling}
\label{sec:dynamic_linear_modeling}
Another method to capture the phonetic dynamics is to use state space models (SSMs), which are similar to hidden Markov models, except that the hidden states in SMMs are continuous \cite{Murphy2012}. Discrete-time dynamic linear models (DLMs) with Gaussian noise are simple versions of SSMs. It is assumed that the state of a process can be summarized by a multivariate variable, which cannot be observed directly at each time step but can be linked to observables using a measurement equation, provided that all the conditional random variables are linear-Gaussian distributed \cite{Roweis1999}. In the last decade, DLMs have been used to characterize acoustic-feature dynamics and track objects in videos \cite{Ma2014,Veeraraghavan2005}. Inspired by these studies, we assume that each utterance can be generated by a DLM-based system, which may be regarded as a sub-system related to an unknown \textit{language production system} in the human brain. This assumption is similar to the acoustic theory of speech production, which assumes that speech production is a linear system comprising a source and a filter \cite{Rabiner1978}. 

Let $\{\mathbf{y}_k\}_{k=1}^K$ be a sequence of $K$ phonetic vectors of an utterance, where $\mathbf{y}_k\in\mathbb{R}^M$ and $M$ denotes the number of phonemes, and $\{\mathbf{x}_k\}_{k=1}^K$ be the corresponding latent variables representing the hidden states of the system, where $\mathbf{x}_k\in\mathbb{R}^d$, $d$ denotes the system complexity, and $d\leq \min (M, K)$. A DLM can be specified by the following two linear equations:
\begin{subequations}
\label{eq:dlm}
\begin{align}
\mathbf{x}_k&=\mathbf{A}\mathbf{x}_{k-1}+\mathbf{w},\quad &\mathbf{w}\sim N(\mathbf{0},\mathbf{Q});\\
\mathbf{y}_k&=\mathbf{C}\mathbf{x}_k+\mathbf{v},\quad &\mathbf{v}\sim N(\mathbf{0},\mathbf{R}).
\end{align}
\end{subequations}
In (\ref{eq:dlm}), $\mathbf{w}\in\mathbb{R}^d$ and $\mathbf{v}\in\mathbb{R}^M$ denote two time-independent noises caused during state evolution and observation generation, respectively. They are white-Gaussian distributed and stochastically independent. In addition to the covariance matrices $\mathbf{Q}$ and $\mathbf{R}$, which can be reduced to identity matrices $\mathbf{I}_d$ and $\mathbf{I}_M$ for simplicity, respectively, the system comprises two operators that must be estimated, i.e., the state transition matrix $\mathbf{A}\in\mathbb{R}^{d\times d}$ and the generative matrix $\mathbf{C}\in\mathbb{R}^{M\times d}$. In this regard, the DLM transforms an utterance into a pair of linear operators, i.e., $\mathbf{A}$ and $\mathbf{C}$, as shown in Fig. \ref{fig:pv_context}. 

To learn the dynamic system, i.e., to identify the parameters $\mathbf{A}$ and $\mathbf{C}$, it is typically assumed that the distribution of the input sequence $\{\mathbf{y}_k\}_{k=1}^K$ is known. Therefore, we can infer the state sequence $\{\mathbf{x}_k\}_{k=1}^K$ and estimate the parameters using the expectation-maximization (EM) algorithm based on the maximum likelihood estimation or maximum a posteriori \cite{Ghahramani1996}. However, owing to its sensitivity to initial parameter estimates, EM does not always yield satisfactory results, as described in \cite[Chapter~18]{Murphy2012}. Therefore, we adopted an alternative approach that estimates the parameters by the reconstruction error minimization of (\ref{eq:dlm}). This approach is known as a subspace method \cite{Katayama1996}, \cite[Chapter~7]{Ljung1999}. We aim to search for a state-space realization that does not require noise modeling. The algorithm used to derive the closed-form solutions of $\mathbf{A}$ and $\mathbf{C}$ is shown in Algorithm \ref{alg:dlm} in the Appendix.

\subsection{Orthogonality Pursuit}
\label{sec:orthogonality_pursuit}
The subspace mentioned in this section is not merely a bag of fixed-size vectors; instead, it has its own topological meaning that must be clarified by the concept of a manifold. A manifold $\mathcal{M}$ is a topological space that is locally homeomorphic to the Euclidean space $\mathbb{R}^D$ for some $D$; this also represents the dimensionality of the manifold. Hence, a bijective mapping from an open set around each point $m\in \mathcal{M}$ to an open set in $\mathbb{R}^D$ exists. A manifold endowed with the differentiable property and a Riemannian metric is called a Riemannian manifold, which extends measures such as lengths and \textit{angles} from familiar Euclidean spaces to a curved space \cite{Edelman1998}. Two of its applicable members, the Stiefel and Grassmann manifolds, with their embeddings in $\mathbb{R}^D$ might help us understand the reality of subspaces and their metrics.

The set of $D\times d$ matrices with orthonormal columns forms the Stiefel manifold $\mathcal{ST}_{D,d}$, where $D\geq d\geq 1$. In the extreme case, $\mathcal{ST}_{D,d}$ results in the group of all the $d\times d$ orthogonal matrices denoted by $\mathcal{O}_d$ if $D=d$. We can express $\mathcal{ST}_{D,d}$ as
\begin{equation}
\label{eq:stiefel}
\mathcal{ST}_{D,d}\vcentcolon=\{\mathbf{S}\in \mathbb{R}^{D\times d}|\mathbf{S}^T\mathbf{S}=\mathbf{I}_d\}.
\end{equation}

Owing to the defined inner product and tangents, $\mathcal{ST}_{D,d}$, which is regarded as an embedded manifold in the Euclidean space $\mathbb{R}^D$, inherits a canonical Riemannian metric in $\mathbb{R}^{D\times d}$ \cite{Edelman1998}. This type of metric enables us to define various geometric notions on the manifold, such as the angle between two curves (or linear subspaces in view of $\mathbb{R}^D$) and the length of a curve. In the following, several methods are proposed to derive a set of orthonormal vectors from stacked vectors or the dynamic linear model of an utterance.

\subsubsection{From a set of stacked vectors to a subspace}
\label{sec:stacked_vector_to_subspace}
Suppose $K\geq D \geq d$, the simplest method to obtain $\mathbf{S}\in\mathbb{R}^{D\times d}$ in (\ref{eq:stiefel}) for an utterance is to seek a set of orthonormal bases that compose the column space, where all contextualized phonetic vectors $\{\mathbf{z}_k\}_{k=1}^K$ defined in Section \ref{sec:vector_stacking} are spanned approximately. This objective can be achieved by solving the following orthogonal linear regression (OLR) problem in a matrix-factorization manner, which involves obtaining the best orthogonal projections $\hat{\mathbf{S}}_{olr}$ while minimizing the reconstruction error \cite{Hamm2008}:
\begin{subequations}
\label{eq:olr}
\begin{align}
\hat{\mathbf{S}}_{olr}=\mathop{\arg\min}_{\mathbf{S}}\|\mathbf{Z}-\mathbf{S}\mathbf{W}\|^2_F\\
\text{subject to}~\mathbf{S}^T\mathbf{S}=\mathbf{I}_d,
\end{align}
\end{subequations}
where $\mathbf{Z}=[\mathbf{z}_1,\dotsc,\mathbf{z}_K]\in\mathbb{R}^{D\times K}$, and $\|\cdot\|_F$ denotes the Frobenius norm. In fact, because $d\leq D$, the unbounded loading matrix $\mathbf{W}\in\mathbb{R}^{d\times K}$ renders (\ref{eq:olr}) an unweighted low-rank approximation problem, which can be solved through truncated SVD, while naturally ensuring the satisfaction of the orthogonality constraint. Therefore, we approximated $\mathbf{Z}$ by $\mathbf{U}\mathbf{\Sigma}\mathbf{V}^T$, where $\mathbf{\Sigma}$ is a $d\times d$ diagonal matrix containing the largest $d$ singular values $\sigma_1,\dotsc,\sigma_d$ in descending order, and $\mathbf{U}$ and $\mathbf{V}$ are matrices with orthonormal columns $\{\mathbf{u}_i\}_{i=1}^d$ and $\{\mathbf{v}_i\}_{i=1}^d$ spanning the column and row spaces of $\mathbf{Z}$, respectively. Finally, we express $\hat{\mathbf{S}}_{olr}=[\mathbf{u}_1,\dotsc,\mathbf{u}_d]$ as the ultimate subspace-based representation of an utterance.

To further filter out some utterance-dependent noise or irrelevant information varying among the contextualized vectors ${\{\mathbf{z}_k\}}$ and within the space spanned by $\mathbf{S}$, we considered the sparsity of the loading matrix $\mathbf{W}$, where each column vector $\mathbf{w}_k$ contains less than $d$ non-zero values \cite{Huang2006}. Therefore, (\ref{eq:olr}) can be reformulated as solving the orthogonal dictionary learning (ODL) problem as follows: 
\begin{subequations}
\label{eq:odl}
\begin{align}
\hat{\mathbf{S}}_{odl}=\mathop{\arg\min}_{\mathbf{S}}\|\mathbf{Z}-\mathbf{S}\mathbf{W}\|^2_F+\lambda^2_{odl}\|\mathbf{W}\|_0 \\
\text{subject to}~\mathbf{S}^T\mathbf{S}=\mathbf{I}_d,
\end{align}
\end{subequations}
where $\|\mathbf{W}\|_0$ denotes the sparsity measure defined as the number of nonzero entries in $\mathbf{W}$, and $\lambda_{odl} >0$ is a scalar regularization parameter that balances the tradeoff between the reconstruction error and sparsity. Because $\mathbf{W}$ is a sparse matrix, only a few column bases in $\mathbf{S}$ contribute to representing some ${\mathbf{z}_k}$ in terms of ${\mathbf{w}_k}$; the remaining column bases contribute to the restoration of unwanted residual information. In fact, if each ${\mathbf{z}_k}$ represents a contextualized sound pattern, a relatively large $\lambda_{odl}$ in (\ref{eq:odl}) guarantees that ${\mathbf{z}_k}$ is linearly compounded from only a small number of basic components. This corresponds analogously to the attribute-based SLR approach, where each canonical phone is tokenized by only two types of articulatory attributes, i.e., manner and place \cite{Siniscalchi2009}.

In contrast to traditional dictionary learning, $\mathbf{S}\in\mathbb{R}^{D\times d}$ in (\ref{eq:odl}) is \textit{not} over-complete because of $d\leq D$, and has an orthogonal constraint on itself. Therefore, we cannot directly follow conventional procedures for ODL \cite{Mairal2009}. Propositions for solving (\ref{eq:odl}) are presented in the Appendix. The detailed procedure is presented in Algorithm \ref{alg:solr}. 

It is noteworthy that the current OLR and ODL-based subspace construction methods cannot address short utterances. This is because the number of phonetic vectors (i.e., the length of a phone sequence) $K$ of an utterance must be greater than or equal to the dimension of the stacked vector, $D$. Because a 30-s utterance in general contains 300--500 phones and the sizes of phone sets used in our phone recognizers range from 45 to 61, the condition is always satisfied. The issue of handling short utterances will be addressed in our future research.

\subsubsection{From a dynamic linear model to a subspace}
\label{sec:dlm_to_subspace}
In \cite{DeCock2002}, Cock and Moor proposed a distance measurement between two dynamic linear models. In this study, we used subspace-based methods to calculate the distance between two DLMs. For the estimates of the state transition matrix $\hat{\mathbf{A}}\in\mathbb{R}^{d\times d}$ and the generative matrix $\hat{\mathbf{C}}\in\mathbb{R}^{M\times d}$ derived through Algorithm \ref{alg:dlm}, the corresponding subspace-based representation is defined as
\begin{equation}
\label{eq:dlm_2}
\hat{\mathbf{S}}_{dlm}\vcentcolon=\left[\hat{\mathbf{C}}^T,(\hat{\mathbf{C}}\hat{\mathbf{A}})^T,\dotsb,(\hat{\mathbf{C}}\hat{\mathbf{A}}^{(n-1)})^T\right]^T,
\end{equation}
where $n$ denotes the context order. Furthermore, $\hat{\mathbf{S}}_{dlm}$ is named the observability matrix \cite{Katayama1996} because it characterizes how a state vector $\mathbf{x}_k$ associated with the $k$th sound pattern in an utterance affects the generation of the $k$th phonetic vector $\mathbf{y}_k$ and the next $n-1$ phonetic vectors $\mathbf{y}_{k+1},\dotsc,\mathbf{y}_{k+n-1}$:
\begin{equation}
\label{eq:dlm_3}
\begin{bmatrix}
\mathbf{y}_k \\
\mathbf{y}_{k+1} \\
\vdots \\
\mathbf{y}_{k+n-1}
\end{bmatrix}
=
\begin{bmatrix}
\hat{\mathbf{C}} \\
\hat{\mathbf{C}}\hat{\mathbf{A}} \\
\vdots \\
\hat{\mathbf{C}}\hat{\mathbf{A}}^{(n-1)}
\end{bmatrix}
\mathbf{x}_k.
\end{equation}

Because we have ensured that $\hat{\mathbf{C}}$ and $\hat{\mathbf{A}}$ are both orthonormal in steps 2 and 4 of Algorithm \ref{alg:dlm}, we obtained $\hat{\mathbf{S}}^T_{dlm}\hat{\mathbf{S}}_{dlm}=\mathbf{I}_d$ and $\mathrm{rank}(\hat{\mathbf{S}}_{dlm})=d$. Consequently, $\hat{\mathbf{S}}_{dlm}$ belongs to a Stiefel manifold, which is necessary for calculating the principal angles between two subspaces. 

\subsection{Summary}
\label{sec:summary}
In summary, three implementations can be performed for subspace construction: OLR (cf. (\ref{eq:olr})), ODL (cf. (\ref{eq:odl})), and DLM (cf. (\ref{eq:dlm_2})). Their goals are summarized in Table \ref{tab:symmary}. For OLR and ODL, an utterance is represented by a subspace characterized by $d$ $D$-dimensional bases that approximately span the $K$ $D$-dimensional stacked phonetic vectors of the utterance. In this case, $d$ determines the amount of information carried in a subspace. For DLM, the phonetic vectors of an utterance are modeled by a dynamic linear system characterized by $d$ $D$-dimensional vectors, where $d$ denotes the complexity of the system represented by a subspace. Although the physical meanings of $d$ in both cases differ, the symbol $d$ reflects a common geometric meaning, i.e., the number of orthogonal bases used to represent an utterance.

In the next section, we present two methods for calculating the distance between two utterances based on their subspace representations.

\begin{table}[t]
\caption{Summary of our proposed methods for subspace construction.}
\vspace{-5pt}
\label{tab:symmary}
\centering
\begin{tabular}{lcc}
\toprule
{\bf Method} & {\bf abbr.} & {\bf Equation} \\
\midrule
{\bf Orthogonal Linear Regression} & OLR & (\ref{eq:olr}) \\
\multicolumn{3}{l}{To minimize the reconstruction errors of the augmented phonetic vectors} \\
\multicolumn{3}{l}{under considerations of the orthogonality and 2-norm regularization.} \\
\midrule
{\bf Orthogonal Dictionary Learning} & ODL & (\ref{eq:odl}) \\
\multicolumn{3}{l}{The same as OLR but with 1-norm regularization.} \\
\midrule
{\bf Dynamic Linear Modeling} & DLM & (\ref{eq:dlm_2}) \\
\multicolumn{3}{l}{To solve the problem of a dynamic linear system by the subspace method.} \\
\bottomrule
\end{tabular}
\vspace{-15pt}
\end{table}

\section{Subspace Learning}
\label{sec:subspace_learning}
To learn from subspaces, the equivalence of subspaces and their mutual similarity measure must be defined and clarified in advance. Because a subspace may have countless matrix-like expressions, the distance between two subspaces cannot be measured by matrix-matrix subtraction.

In addition to $\mathbf{S}\in\mathcal{ST}_{D,d}$, we assume that $\mathbf{S}\in\mathcal{G}_{D,d}$, where $\mathcal{G}_{D,d}$ denotes the Grassmann manifold defined by
\begin{equation}
\label{eq:grassmann}
\mathcal{G}_{D,d}\vcentcolon=\mathcal{ST}_{D,d}/\mathcal{O}_d.
\end{equation}
Here, $\mathcal{G}_{D,d}$ is the quotient of group $\mathcal{ST}_{D,d}$ modulo its subgroup $\mathcal{O}_d$. Therefore, we can identify an abstract \textit{point} (subspace) $\mathbf{S}\in\mathbb{R}^{D\times d}$ with orthogonal columns on $\mathcal{G}_{D,d}$ as an equivalent class under the orthogonal transform of $\mathcal{ST}_{D,d}$. Two subspaces $\mathbf{S}_1,\mathbf{S}_2\in \mathbb{R}^{D\times d}$ are equivalent if an orthogonal matrix $\mathbf{Q}\in\mathcal{O}_d$ exists such that $\mathbf{S}_1=\mathbf{S}_2\mathbf{Q}$ \cite{Edelman1998}.

In the following, we present the definition of the principal angle (or canonical correlation) between two subspaces, the corresponding distance function and kernel function applicable to SVMs, and the subspace-based weights for SNNs.

\subsection{Principal Angles, Projection Kernel, and SVMs}
\label{sec:svms}
For two subspaces $\mathbf{S}_1\in \mathbb{R}^{D\times d}$ and $\mathbf{S}_2\in \mathbb{R}^{D\times d}$ ($D\geq d$), the $d$ principal (or minimal) angles between them are recursively and uniquely defined as ${\theta_1,\dotsc,\theta_d}$ in $[0, \pi /2]$ satisfying
\begin{subequations}
\label{eq:pa}
\begin{align}
\cos{\theta_k}=\max_{\substack{\|\mathbf{a}\|_2=1 \\ \mathbf{a}\in\mathrm{span}(\mathbf{S}_1)}}\max_{\substack{\|\mathbf{b}\|_2=1 \\ \mathbf{b}\in\mathrm{span}(\mathbf{S}_2)}}\mathbf{a}^T\mathbf{b}=\mathbf{a}_k^T\mathbf{b}_k,\\
\text{subject to}~\mathbf{a}^T\mathbf{a}_i=\mathbf{b}^T\mathbf{b}_i=0, i=1,\dotsc,d-1.
\end{align}	
\end{subequations}
Because the columns of $\mathbf{S}_1$ and $\mathbf{S}_2$ form unitary bases of the two subspaces, $\cos{\theta_k}$ can be derived by performing SVD on $\mathbf{S}_1^T\mathbf{S}_2$, i.e., $\mathbf{S}_1^T\mathbf{S}_2=\mathbf{U}\mathbf{\Sigma}\mathbf{V}^T$, where $\mathbf{U}=\{\mathbf{a}_1,\dotsc,\mathbf{a}_d\}$, $\mathbf{V}=\{\mathbf{b}_1,\dotsc,\mathbf{b}_d\}$, and $\mathbf{\Sigma}=\mathrm{diag}(\cos{\theta_1},\dotsc,\cos{\theta_d})$ \cite{Bjorck1973,Golub2012}.
The cosine-based similarity between $\mathbf{S}_1$ and $\mathbf{S}_2$ can be defined as
\begin{equation}
\label{eq:sim}
\mathrm{sim}(\mathbf{S}_1,\mathbf{S}_2)\vcentcolon=\sum_{i=1}^d\cos^2\theta_i.
\end{equation}
Because the Euclidean distance between two vectors can be calculated by the dot product, i.e., $\|\mathbf{u}-\mathbf{v}\|^2=\mathbf{u}^T\mathbf{u}+\mathbf{v}^T\mathbf{v}-2\mathbf{u}^T\mathbf{v}$, we can represent (\ref{eq:sim}) as a distance function: 
\begin{equation}
\label{eq:dist}
\begin{split}
\mathrm{dist}^2(\mathbf{S}_1,\mathbf{S}_2)&\vcentcolon=\mathrm{sim}(\mathbf{S}_1,\mathbf{S}_1)+\mathrm{sim}(\mathbf{S}_2,\mathbf{S}_2)-2\mathrm{sim}(\mathbf{S}_1,\mathbf{S}_2) \\
&=2d-2\|\mathbf{S}_1^T\mathbf{S}_2\|_F^2 \\
&=\|\mathbf{S}_1\mathbf{S}_1^T-\mathbf{S}_2\mathbf{S}_2^T\|_F^2 \\
&=\|\Psi(\mathbf{S}_1)-\Psi(\mathbf{S}_2\|_F^2,
\end{split}
\raisetag{1\baselineskip}
\end{equation}
where $\Psi: \mathrm{span}(\mathbf{S})\mapsto  \mathbf{S}\mathbf{S}^T$ maps all subspaces from $\mathbb{R}^D$ to a higher-dimensional feature space $\mathbb{R}^{D\times D}$ \cite{Chikuse2003}. For any two subspaces, $\Psi$ satisfies $\mathrm{span}(\mathbf{S}_1)=\mathrm{span}(\mathbf{S}_2)\iff \Psi(\mathbf{S}_1)=\Psi(\mathbf{S}_2)$ \cite{Hamm2008}. Subsequently, $\mathrm{trace}((\mathbf{S}_a\mathbf{S}_a^T)(\mathbf{S}_b\mathbf{S}_b^T))$, as described in \cite[Section~9.2]{Hoffman1971}, can be adopted as the inner product of $\Psi(\mathbf{S}_1)$ and $\Psi(\mathbf{S}_2)$.  Using the kernel trick \cite{Scholkopf2000}, the subspace-based kernel function can be obtained as
\begin{equation}
\label{eq:kernel}
\begin{split}
\mathrm{kern}(\mathbf{S}_1,\mathbf{S}_2)&=\langle \Psi(\mathbf{S}_1),\Psi(\mathbf{S}_2)\rangle=\mathrm{trace}((\mathbf{S}_1\mathbf{S}_1^T)(\mathbf{S}_2\mathbf{S}_2^T)) \\
&=\|\mathbf{S}_1^T\mathbf{S}_2\|_F^2.
\end{split}
\raisetag{0.7\baselineskip}
\end{equation}
It is easy to verify that $\mathrm{kern}(\mathbf{S}_1,\mathbf{S}_2)=\mathrm{kern}(\mathbf{S}_1\mathbf{Q}_1,\mathbf{S}_2\mathbf{Q}_2)$, where $\mathbf{Q}_1,\mathbf{Q}_2\in\mathcal{O}_d$. To see the positive definiteness of the kernel, i.e., it can give rise to a positive Gram matrix, we have 
\begin{equation}
\label{eq:pd}
\begin{aligned}
&\sum_{i,j}\nolimits{c_i c_j\|\mathbf{S}_i^T\mathbf{S}_j\|_F^2}=\sum_{i,j}\nolimits{c_i c_j\mathrm{trace}(\mathbf{S}_i\mathbf{S}_i^T\mathbf{S}_j\mathbf{S}_j^T)} \\
&=\sum_{i,j}\nolimits{\mathrm{trace}((c_i\mathbf{S}_i\mathbf{S}_i^T)(c_j\mathbf{S}_j\mathbf{S}_j^T))} \\
&=\mathrm{trace}((\sum_{i}\nolimits{c_i\mathbf{S}_i\mathbf{S}_i^T})^2)=\|\sum_{i}\nolimits{c_i \mathbf{S}_i\mathbf{S}_i^T}\|_F^2\geq 0.
\end{aligned}
%\raisetag{0.7\baselineskip}
\end{equation}
Therefore, (\ref{eq:kernel}), named \textit{the Projection kernel} \cite{Hamm2008}, enables us to build a precomputed kernel matrix for valid SVM learning for SLR. We refer readers to \cite[Chapters~2~and~7.4]{Scholkopf2001} for details.

\subsection{SNNs}
\label{sec:snns}
The SNNs for the SLR, as shown in Fig. \ref{fig:snns}, transform the subspace inputs, contributed by a set of phone recognizers, into a target label for each utterance. There are two hidden layers. One processes subspace-based inputs to generate subspace-aware scores (cf. the kernel layer in Fig. \ref{fig:snns}), whereas the other involves multilayer perceptrons (MLPs) followed by an output layer with a softmax activation function to transform the subspace-aware scores into the decision scores for classification. The most significant differences between SNNs and regular DNNs are as follows:

\begin{figure}[!t]
\centering
\includegraphics[width=0.48\textwidth]{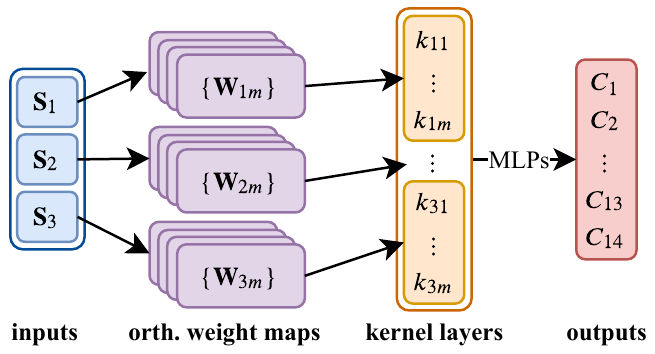}
\vspace{-10pt}
\caption{The subspace-based neural networks (SNNs), where 3 subspace inputs are associated with 3 BUT phone recognizers respectively, the output layer is associated with 14 target languages, and \textit{m} orthonormal weight maps per input are for kernel computation.}
\label{fig:snns}
\vspace{-15pt}
\end{figure}

\textbf{Inputs}. The input layer of SNNs simultaneously accepts multiple subspace inputs $\mathbf{S}_{1},\dotsc,\mathbf{S}_{L}$ of an utterance, each from one of the $L$ parallel phone recognizers. The ranks (widths) of the inputs can be varied. For example, if the utterance is represented by two subspace inputs, $\mathbf{S}_{1}\in\mathbb{R}^{D_1\times d_1}$ and $\mathbf{S}_{2}\in\mathbb{R}^{D_2\times d_2}$, they are allowable inputs for the following weight computation even when $D_1\neq D_2$.

\textbf{Orthonormal Weight Maps}. The second layer of SNNs is a kernel function $f_{lm}:\mathbb{R}^{D_l\times d_l}\rightarrow\mathbb{R}$, which, in matrix notation, can be expressed as
\begin{subequations}
\label{eq:weight}
\begin{align}
f_{lm}(\mathbf{S}_{l})=k_{lm}=\|\mathbf{W}_{lm}^T\mathbf{S}_{l}\|_F^2\\
\text{subject to}~\mathbf{W}_{lm}^T\mathbf{W}_{lm}=\mathbf{I}_{d'_l},
\end{align}
\end{subequations}
where $\mathbf{W}_{lm}\in\mathbb{R}^{D_l\times d'_l}$ denotes the $m$-th orthonormal weight map or kernel subspace for the $l$-th subspace input with $d'_l$ as its rank, and $\|\cdot\|_F$ denotes the Frobenius norm used to measure the similarity between a \textit{sample} subspace $\mathbf{S}_{il}\in\mathcal{G}_{D_l,d_l}$ and a \textit{reference} subspace $\mathbf{W}_{lm}\in\mathcal{G}_{D_l,d'_l}$. In fact, (\ref{eq:sim}) is not confined to two subspaces lying on the same Grassmann manifold \cite{Golub2012}. The similarity between two subspaces $\mathbf{W}\in\mathcal{G}_{D,p}$ and $\mathbf{S}\in\mathcal{G}_{D,q}$ can be measured based on their principal angles by $\sum_{i=1}^q\cos^2\theta_i=\|\mathbf{W}^T\mathbf{S}\|_F^2$ if $p\geq q$. Moreover, it can be easily verified that two \textit{sample} subspaces will have the same kernel score $k_{lm}$ when they are equivalent in the Grassmannian sense.

\textbf{Optimization and Regularization}. To train SNNs, a back-propagation process is implemented from the output layer down through the whole SNNs to adjust all parameters. The gradient of each weight can be easily derived by partial differentiation on the cross-entropy cost function, so that the model can be iteratively updated by using an optimizer based on gradient descent. Equation (\ref{eq:weight}a) is differentiable with respect to $\mathbf{W}_{lm}$, and the derivative is 
\begin{equation}
\begin{aligned}
&\frac{\partial f_{lm}(\mathbf{S}_{il})}{\partial\mathbf{W}_{lm}}=\frac{\partial\|\mathbf{W}_{lm}^T\mathbf{S}_{il}\|_F^2}{\partial\mathbf{W}_{lm}} \\
&=\frac{\partial \mathrm{trace}(\mathbf{W}_{lm}^T\mathbf{S}_{il}\mathbf{S}_{il}^T\mathbf{W}_{lm})}{\partial\mathbf{W}_{lm}}=2\mathbf{S}_{il}\mathbf{S}_{il}^T\mathbf{W}_{lm}.
\end{aligned}
\end{equation}
However, the constraint term in (\ref{eq:weight}b) renders the parameter estimation intractable. To alleviate this problem, an alternative method is to replace (\ref{eq:weight}b) with the following regularizer that allows us to apply penalties on $\mathbf{W}_{lm}$ as follows:
\begin{equation}
\label{eq:reg}
\lambda\|\mathbf{W}_{lm}^T\mathbf{W}_{lm} - \mathbf{I}_{d'_l}\|_F^2,
\end{equation}
where the tunable parameter $\lambda$ controls the degree of orthonormality of $\mathbf{W}_{lm}$. These penalties are further incorporated in the final categorical cross-entropy loss (i.e., negative log-likelihood) function with a derivative with respect to $\mathbf{W}_{lm}$,
\begin{equation}
\label{eq:reg_grad}
2\lambda\mathbf{W}_{lm}(\mathbf{W}_{lm}^T\mathbf{W}_{lm} - \mathbf{I}_{d'_l}),
\end{equation}
while performing optimization using back propagation.

\begin{table}[t]
\setlength{\tabcolsep}{4pt}
\caption{The number of utterances in different tasks with respect to different languages. TR and TEST denote the training and test data, respectively.}
\vspace{-5pt}
\label{tab:stats_lre}
\centering
\begin{tabular}{lrrrrrr}
\toprule
{\bf Task} & \multicolumn{2}{c}{\bf LRE-03} & \multicolumn{2}{c}{\bf LRE-07} & \multirow{2}{*}{\bf Total} \\
\cmidrule(lr){2-3} \cmidrule(lr){4-5}
\bf Language & \multicolumn{1}{c}{\bf TR} & \multicolumn{1}{c}{\bf TEST (30 s)} & \multicolumn{1}{c}{\bf TR} & \multicolumn{1}{c}{\bf TEST (30 s)} \\
\midrule
Arabic & 309 & 80 & 629 & 80 & 1,098 \\
\rowcolor{lightgray} Bengali & & & 240 & 80 & 320 \\
Chinese & 625 & 80 & 5,549 & 398 & 6,652 \\
\rowcolor{lightgray} English & 1,276 & 240 & 6,684 & 240 & 8,440 \\
Farsi & 318 & 80 & 1,253 & 80 & 1,731 \\
\rowcolor{lightgray} French & 314 & 80 & - & 80 & 474 \\
German & 318 & 80 & 646 & 80 & 1,124 \\
\rowcolor{lightgray} Hindustani & 309 & 80 & 2,956 & 240 & 3,585 \\
Indonesian & & & - & 80 & 80 \\
\rowcolor{lightgray} Italian & & & - & 80 & 80 \\
Japanese & 315 & 160 & 1,702 & 80 & 2,257 \\
\rowcolor{lightgray} Korean & 312 & 80 & 2,014 & 80 & 2,486 \\
Punjabi & & & - & 32 & 32 \\
\rowcolor{lightgray} Russian & - & 80 & 1,323 & 160 & 1,563 \\
Spanish & 611 & 80 & 2,763 & 240 & 3,694 \\
\rowcolor{lightgray} Tagalog & & & - & 80 & 80 \\
Tamil & 297 & 80 & 823 & 160 & 1,360 \\
\rowcolor{lightgray} Thai & & & 515 & 80 & 595 \\
Vietnamese & 309 & 80 & 1,069 & 160 & 1,618 \\
\midrule
{\bf Total} & 5,313 & 1,280 & 28,166 & 2,510 & 37,269 \\
\midrule
$\displaystyle\frac{\textbf{Positives}}{\textbf{Trials}}$ & & $\displaystyle\frac{1,200}{15,360}$ & & $\displaystyle\frac{2,158}{35,140}$ & \\
\bottomrule
\end{tabular}
\vspace{-15pt}
\end{table}

\section{Experiment Results and Discussion}
\label{sec:experiments}
Our proposed methods were evaluated on two tasks: NIST LRE 2007 (LRE-07) \cite{NIST2007} for spoken language verification (SLV) and NIST LRE 2009 (LRE-09) \cite{NIST2009} for dialect/accent identification. 

The first task pertains to the \textit{General LR} test of LRE-07 conducted on 30-s 8k-sampled telephony utterances. Nineteen languages were tested, including 14 target languages and 5 out-of-set languages (French, Indonesian, Italian, Punjabi, and Tagalog) that were not the target languages of each trial and lacked corresponding training data. Four target languages (Chinese, English, Hindustani, and Spanish) contained sublanguages or dialects. To learn the phonotactic models and backend classifiers, we drew training utterances that were longer than 10 s from NIST LRE 1996 (LRE-96, LDC2006S31), 2003 (LRE-03, LDC2006S31), 2005 (LDC2008S05), 2009 (LDC2014S06), and the supplemental set provided by LRE-07 (LDC2009S05). We used LRE-03 as the validation set to determine the optimal representation setting and model hyperparameters. According to the LRE-03 evaluation plan \cite{NIST2003}, the test set contains 12 target languages and one out-of-set language (Russian); the training data were only from LRE-96. The number of utterances and the statistics regarding trials in the LRE-07 (for evaluation) and LRE-03 (for validation) tasks are listed in Table \ref{tab:stats_lre}.

The second task involves 8 dialect/accent pairs, as shown in Table \ref{tab:stats_daid}, for evaluating how a system can distinguish cognate or mutually intelligible languages. Each dialect/accent pair, where only 10-s and 30-s recorded telephony utterances were selected from LRE-09, was independently evaluated by binary closed-set classification using stratified five-fold cross-validation.

\subsection{Phone Recognition}
\label{sec:phone_recognition}
We used the well-known BUT phone recognizers\footnote{\url{https://speech.fit.vutbr.cz/software/phoneme-recognizer-based-long-temporal-context}} developed by the Brno University of Technology for Czech (CZ), Hungarian (HU) and Russian (RU) \cite{Schwarz2008}. The numbers of sound units in the BUT decoders for CZ, HU, and RU were 46, 62, and 53, respectively. Each set includes three nonphonetic units, i.e., /int/ (intermittent noise), /pau/ (short pause), and /spk/ (nonspeech speaker noise). These phone decoders use a three-state model per unit, which means that three posterior probabilities per unit are provided at each frame. Therefore, to form the phonetic vectors for our proposed subspace construction, as shown in Fig. \ref{fig:pv_context}, a single posterior probability was computed for each unit by adding the posterior probabilities of all the states in the corresponding model.

\newcolumntype{R}[1]{>{\raggedleft}m{#1}}

\begin{table}[t]
\caption{The number of utterances used in accent/dialect identification with respect to 8 pairs of languages.}
\vspace{-5pt}
\label{tab:stats_daid}
\centering
\begin{tabular}{llr>{\hspace{-10pt}}rc}
\toprule
{\bf Dialect or Accent Pair} & {\bf abbr.} & \multicolumn{2}{c}{\bf \# Utt.} & {\bf Total} \\
\midrule
Mandarin / Cantonese & Man / Can & 2,006 / & 730 & 2,736 \\
\rowcolor{lightgray} Portuguese / Spanish & Por / Spa & 794 / & 770 & 1,564 \\
Haitian Creole / French & Cre / Fre & 646 / & 790 & 1,436 \\
\rowcolor{lightgray} Russian / Ukrainian & Rus / Ukr & 1,003 / & 776 & 1,779 \\
Hindi / Urdu & Hin / Urd & 1,285 / & 756 & 2,041 \\
\rowcolor{lightgray} Dari / Farsi & Dar / Far & 776 / & 775 & 1,551 \\
Bosnian / Croatian & Bos / Cro & 710 / & 752 & 1,462 \\
\rowcolor{lightgray} American Eng. / Indian Eng. & Ame / Ind & 1,759 / & 1,113 & 2,872 \\
\bottomrule
\end{tabular}
\vspace{-15pt}
\end{table}

\begin{figure*}
\centering
\includegraphics[width=0.99\textwidth]{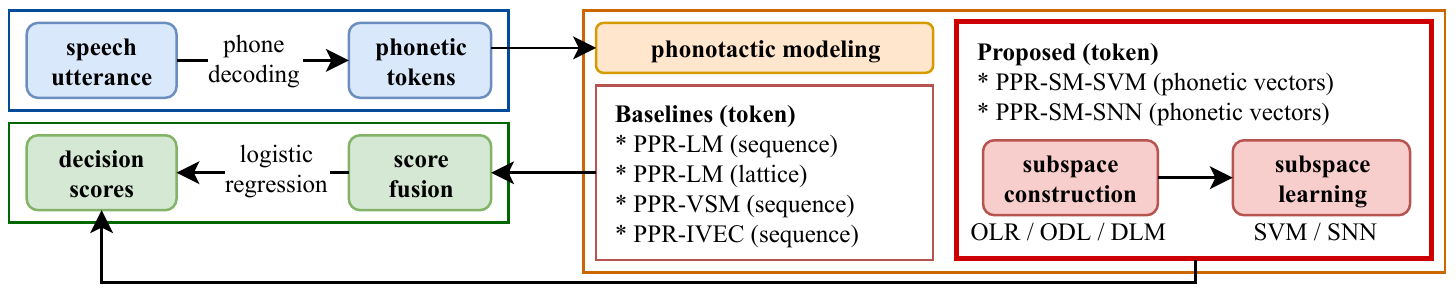}
\vspace{-10pt}
\caption{The experiment flowchart, where the orange box with a dashed bolder shows the procedure of phonotactic modeling, inside which the red rectangle expresses two key steps of our proposed framework based on subspace representation and learning.}
\label{fig:exp_flows}
\vspace{-15pt}
\end{figure*}

\subsection{Experiment Flow and Control}
\label{sec:experiment_setup}
As depicted in Fig. \ref{fig:exp_flows}, parallel phone recognition language modeling (PPR-LM) \cite{Li2013}, parallel phone recognition vector space modeling (PPR-VSM) \cite{Li2013}, and parallel phone recognition subspace multinomial modeling (PPR-IVEC) were implemented as the baselines, where each utterance was represented by multiple multinomial distributions, high-dimensional vectors, and lower-dimensional vectors, respectively. In addition to the sequence-based PPR-LM, we implemented lattice-based PPR-LM \cite{Gauvain2004}, where phone \textit{n}-gram statistics were derived from the top-five decoded phone sequences, and the average numbers of nodes and edges of a phone lattice were 3,700 and 15,000, respectively. For PPR-VSM, three phone-recognizer-dependent 350-dimensional vectors, each of which was derived through phone recognition, \textit{n}-gram counts, TF-IDF weights, and LSA, were evaluated by phone-recognizer-dependent target-dependent one-vs-rest SVMs with a linear kernel \cite{Li2007}. Instead of using NN-based backends such as MLPs, we used SVMs in PPR-VSM because we did not think that MLPs will benefit much from the deterministic vectorial representation. For PPR-IVEC, we adopted the algorithm\footnote{\url{https://github.com/skesiraju/smm}} proposed by Kesiraju \textit{et al.} to represent each utterance as an i-vector using its \textit{n}-gram statistics, where the Laplace prior enforced sparsity in the basis matrix \cite{Kesiraju2016}. Most parameters were the same as those in \cite{Kesiraju2016}, except for the dimension of the i-vector of an utterance, which was 900.

Our proposed method, named parallel phone recognition subspace modeling (PPR-SM), uses two classifiers for subspace learning, i.e., the SVMs and SNNs. The corresponding systems are denoted as PPR-SM-SVM and PPR-SM-SNN, respectively. The PPR-SM-SNN, as shown in Fig. \ref{fig:snns} and (\ref{eq:weight}), is mainly composed of orthonormal weight maps and MLPs. Here, the first set of parameters was initially drawn from the $O(D_l)$ Haar distribution to randomly generate a set of orthonormal matrices \cite{Mezzadri2006}\footnote{For the sake of performance, we did not use Saxe's method \cite{Saxe2014}, which was recently proposed to fill the weights with a (semi) orthogonal matrix and implemented using existing NN-based toolkits, such as Keras and PyTorch.}, and the parameters of the MLPs were normally initialized by the Glorot process \cite{Glorot2010}. The adaptive gradient algorithm for updating the PPR-SM-SNN parameters was Adam \cite{Kingma2015}, where default parameters were used, except that the learning rate was initially fixed at $10^{-3}$. Moreover, a drop-based learning rate schedule was practiced by reducing the learning rate by half every 10 epochs. The mini-batch size and the maximum number of training epochs were 24 and 200, respectively.

Similar to PPR-VSM, PPR-SM-SVM not only specifies the Projection kernel in (\ref{eq:kernel}) on SVMs, but also extends our previous versions in \cite{Shih2012,Lee2013}, where only one phone recognizer developed for a universal phone set is used. For PPR-LM, PPR-VSM, and PPR-SM-SVM, because three BUT phone recognizers and 14 target languages were used in the SLV task, $3\times 14=52$ decision scores contributed by target-dependent LMs and support vectors, respectively, were derived for each test utterance. To combine the 52 scores for one target language, a systematic method based on multiclass logistic regression was implemented \cite{Li2013}. For a fair comparison, the MLPs in the structure of the PPR-SM-SNN (cf. Fig. \ref{fig:snns}) were implemented with no hidden layers, which can be reegarded as a simplified version of logistic regression.

As for subspace construction, three implementations were performed, i.e., OLR, ODL, and DLM, according to (\ref{eq:olr}), (\ref{eq:odl}), and (\ref{eq:dlm_2}). Their goals are summarized in Table \ref{tab:symmary}. For ODL, the regularization parameter $\lambda_{odl}$ and the number of iterations $J$ in Algorithm \ref{alg:solr} were empirically set to $10^{-4}$ and 50, respectively.

\begin{figure*}
\centering
\includegraphics[width=1.0\textwidth]{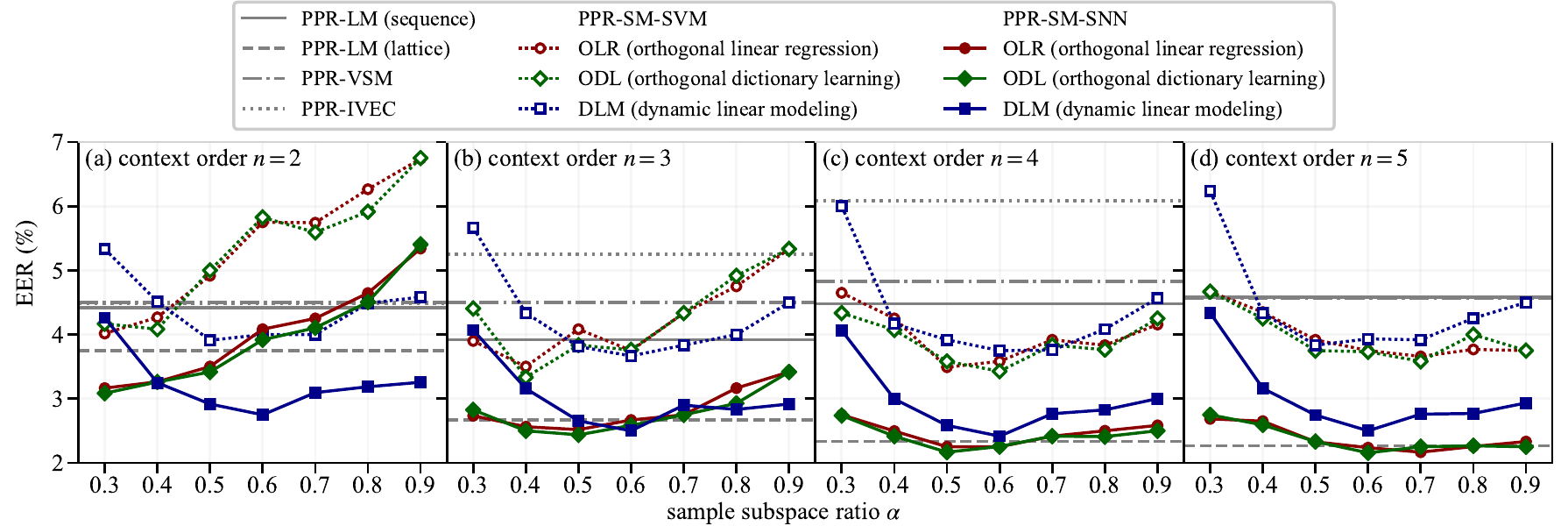}
\vspace{-20pt}
\caption{EERs on LRE-03 for baselines and two kinds of PPR-SM-based backends with respect to various context orders, sample subspace ratios, and three methods for subspace construction: OLR (\ref{eq:olr}), ODL (\ref{eq:odl}), and DLM (\ref{eq:dlm_2}).}
\vspace{-15pt}
\label{fig:lre03_oracle}
\end{figure*}

\subsection{Oracle Experiment on LRE-03}
\label{sec:experiment_lre03}
We conducted oracle experiments on LRE-03 to determine the optimal parameters of the phonotactic models, including the penalty factor in SVMs, the regularization strength in logistic regression, and the subspace dimensionality $d$ in (\ref{eq:olr}), (\ref{eq:odl}), (\ref{eq:dlm_2}), and (\ref{eq:weight}). The rank of the sample subspace $d_{sl}$ was controlled by the sample subspace ratio $\alpha \in (0, 1)$ as $d_{sl}=\max(\floor{\alpha M_l}, 2)$, where $M_l$ denotes the number of phonemes in the $l$-th phone recognizer. The rank of the kernel (reference) subspace $d_{kl}$ was controlled by the reference subspace ratio $\beta\in [0.5, 1.5]$ as $d_{kl}=\max(\floor{\beta d_{sl}}, 2)$. Furthermore, for the PPR-SM-SNN, other parameters, such as the orthogonality factor $\lambda$ in (\ref{eq:reg}) and the number of orthonormal weight maps $m$, were tuned within a heuristic range through an exhaustive grid search, without learning rate scheduling. It is noteworthy that $\beta$ was always set to 1 to satisfy (\ref{eq:pd}) for PPR-SM-SVM.

\begin{figure}[t]
\centering
\includegraphics[width=0.49\textwidth]{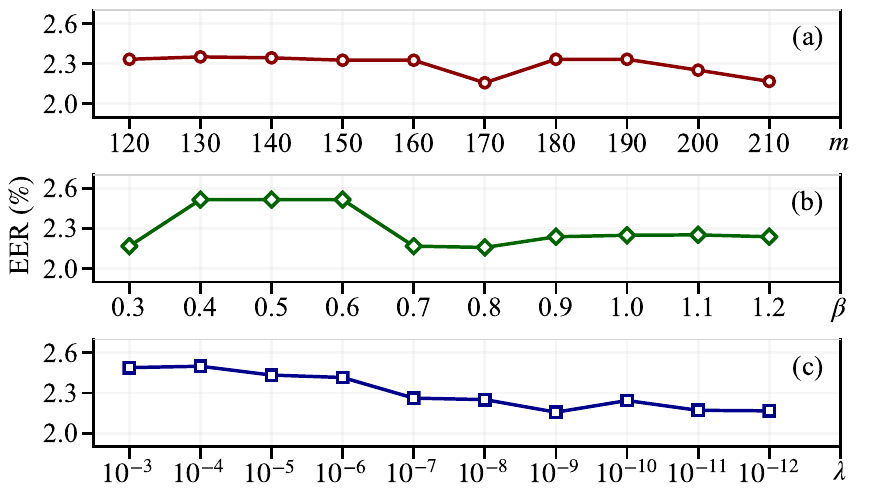}
\vspace{-20pt}
\caption{EERs on LRE-03 for PPR-SM-SNN in the case of ODL/5/0.6 with respect to various values of $m$, $\beta$, and $\lambda$.}
\label{fig:lre03_snn_params}
\vspace{-15pt}
\end{figure}

\subsubsection{Effect of sample subspace ratio}
\label{sec:experiment_subspace_ratio}
Fig. \ref{fig:lre03_oracle} shows the equal error rates (EERs) with respect to $\alpha$ and the context order $n$. Each point (for PPR-SM-SVM and PPR-SM-SNN) and each line (for the baseline methods) represent the best result from experiments with some possible parameter combinations. The parameter $\alpha$ represents the degree that a subspace can describe a set of phonotactic observations. Although a larger $\alpha$ enables a subspace to retain more phonotactic information, it might yield subtle effects caused by unwanted structural noises owing to the deficiency of phone recognizers. The context order $n$ represents the length of the phonotactic constraint used for phonotactic modeling. Although $n=3$ (trigram) has been shown an appropriate length in some phonotactic learning simulation systems \cite{Hayes2008}, we changed $n$ to investigate how these systems achieved their best performances with more or less phonotactic information. The four subplots in Fig. \ref{fig:lre03_oracle} show the manner in which the context order $n$ affects the performance. Unlike PPR-VSM, sequence-based PPR-LM, and PPR-IVEC, where a higher context order may cause overfitting thereby increasing the EERs, the PPR-SM-SNN and lattice-based PPR-LM can utilize the information provided by longer phonotactic constraints. For PPR-SM-SVM, increasing $n$ did not improve performance (the only exception is increasing $n$ from 2 to 3).

\begin{figure}[t]
\centering
\includegraphics[width=0.49\textwidth]{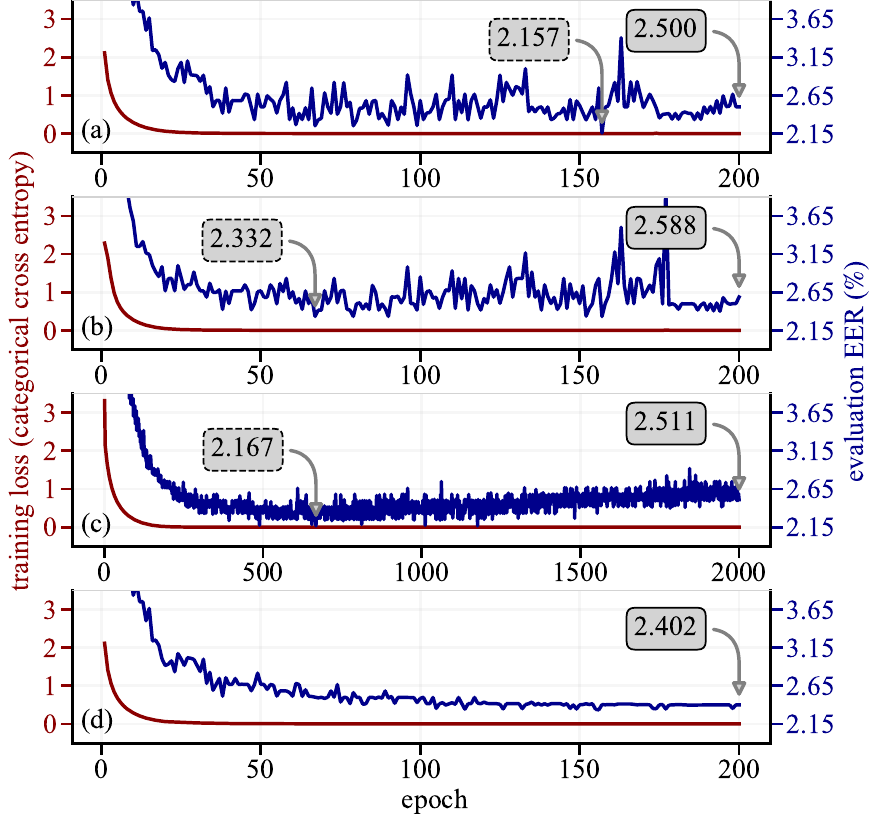}
\vspace{-20pt}
\caption{
The variation of training loss and evaluation EER with respect to the number of training epochs on LRE-03 for PPR-SM-SNN in the case of ODL/5/0.6. The weight map initializer in (a) is orthogonal, while in (b) the Glorot normal initializer is used. Inherited from (a), the learning rate in (c) is reduced from $10^{-4}$ to $10^{-5}$, while in (d), the drop-based learning rate schedule is adopted. The boxes with dashed edges denote the best evaluation EER, while those with solid edges denote the EER after the last epoch.}
\label{fig:lre03_snn_curve}
\vspace{-15pt}
\end{figure}

When focusing on subspace construction, we observed from Fig. \ref{fig:lre03_oracle} that the results of OLR and ODL were similar and better than those of DLM when $n \geq 4$. For DLM, when $\alpha \le 0.6$ (implying a smaller $d$ in Algorithm \ref{alg:dlm}), significant amounts of irrelevant noise and phonetic information might be lost when estimating $\mathbf{C}$ and $\mathbf{X}_{1:K}$ in Algorithm \ref{alg:dlm} and $\mathbf{A}$ in Algorithm \ref{alg:opp}. Moreover, the results show that an increase in context order is not necessarily good for DLM. One possible reason is that, as a type of first-order Markovian dynamics shown in (\ref{eq:dlm}), DLM has a limited ability to explicitly $simulate$ phonotactic constraints that are longer than 3 in length (i.e., $n \ge 3$).

Regarding subspace learning, we discovered that regardless of the subspace construction method used, the PPR-SM-SNN is almost always better than PPR-SM-SVM and is slightly better than lattice-based PPR-LM at certain $\alpha$ values. When the amount of training utterances is relatively large, e.g., exceeding 5,000, as shown in Table \ref{tab:stats_lre}, the PPR-SM-SNN is superior to PPR-SM-SVM. Another limitation of PPR-SM-SVM is the kernel selection \cite{Burges1998}. Only two valid metrics exist on the Grassmann manifold, i.e., the Projection kernel and the Binet--Cauchy kernel, which were studied in \cite{Hamm2008}. The former is more useful for our preliminary experiments. Better kernels for the SLR task will be investigated in future studies.

\subsubsection{Effect of other parameters in PPR-SM-SNN}
\label{sec:experiment_parameter_snn}
Next, we discuss the effects of three key parameters, i.e., $m$ in (\ref{eq:weight}), $\beta$, and $\lambda$ in (\ref{eq:reg}), on the performance of the PPR-SM-SNN. We investigated the best case shown in Fig. \ref{fig:lre03_oracle}, where the subspace construction was based on ODL with $n=5$ and $\alpha=0.6$ (ODL/5/0.6 for short), and the parameters $m$, $\beta$, and $\lambda$ were 170, 0.8, and $10^{-9}$, respectively. In the experiments, each parameter was evaluated separately with the remaining parameters fixed at the best settings described earlier. Based on Fig. \ref{fig:lre03_snn_params}, the following observations were obtained.

\begin{enumerate}[wide,label=\textbf{\roman*)},labelwidth=!,labelindent=1pt]
\setlength{\itemsep}{2pt}
\item
In Fig. \ref{fig:lre03_snn_params}(a), the best number of weight maps $m$ was 170, which makes SNNs a \textit{wide} neural network. 
\item
In Fig. \ref{fig:lre03_snn_params}(b), the best $\beta$ was 0.8 when $\alpha=0.6$, and the corresponding input widths $\{d_{sl}\}_{l=1}^3$ and weight map sizes $\{d_{kl}\}_{l=1}^3$ were [27, 37, 31]\footnote{$d_{sl}=\alpha M_l$, so $[27, 37, 31] \approx 0.6 \times [46, 62, 53]$.} and [21, 29, 24]\footnote{$d_{kl}=\beta d_{sl}$, so $[21, 29, 24] \approx 0.8 \times [27, 37, 31]$.}, respectively, whereas the numbers of phonemes $\{M_l\}_{l=1}^3$ for the BUT phone recognizers were [46, 62, 53]. The result shows that for PPR-SM-SNN, $d_{sl}$ does not have to be equal to $d_{kl}$. However, when $\beta=1$, under the same sample subspace specification, the EER of PPR-SM-SNN was 2.248\%, which was still much lower than that of PPR-SM-SVM (3.732\%).
\item
Although the lowest EER occurred when $\lambda$ was fairly small ($10^{-9}$) , and an almost negative correlation between the orthogonality factor $\lambda$ and EER existed, as shown in Fig. \ref{fig:lre03_snn_params}(c), this does not mean that the orthogonality of the weight maps is less important than expected. We can also observe that when $\lambda$ becomes smaller ($\leq 10^{-10}$), the EER will increase. This result demonstrates that even if the appropriate value of $\lambda$ is small, it still has its role. In general, the negative gradients of commonly used regularizers, such as the $l^2$-norm, always \textit{equi-dimensionally} point toward $\mathbf{0}$. They may not compromise the classification performance but will avoid overfitting the data. Furthermore, this does not mean that the orthonormal initialization of weight maps described in Section \ref{sec:snns} is unnecessary. This issue will be addressed next.
\end{enumerate}

\begin{table*}[ht!]
\setlength{\tabcolsep}{7pt}
\caption{EERs and $C_{avg}$'s on LRE-07 for different methods with our own implementations. The lowest values in each column are in bold face. The column ``rel. \%" shows the relative reduction achieved by PPR-SM-SNN (ODL) over the compared method computed based on their respective best rates among all context orders.}
\vspace{-5pt}
\label{tab:lre07_restlts}
\centering
\begin{tabular}{l|r@{\hskip10pt}r@{\hskip10pt}r@{\hskip10pt}r@{\hskip10pt}r@{\hskip10pt}r|r@{\hskip10pt}r@{\hskip10pt}r@{\hskip10pt}r@{\hskip10pt}r@{\hskip10pt}r}
\toprule
{\bf Metric} & \multicolumn{6}{c}{\bf Equal Error Rate (EER \%)} & \multicolumn{6}{c}{\bf Average Cost ($C_{avg}$)} \\
\cmidrule(lr){2-7} \cmidrule(lr){8-13}
{\bf Context Order $n$ (\textit{n}-gram}) & \bf 2 & \bf 3 & \bf 4 & \bf 5 & \bf rel. \% & \bf p-value & \bf 2 & \bf 3 & \bf 4 & \bf 5 & \bf rel. \% & \bf p-value \\
\midrule
\textbf{PPR-LM (seq.)} \cite{Yan1995} & 10.844 & 10.751 & 9.880 & 10.009 & 52.15 & $<$ 0.001 & 0.152 & 0.151 & 0.132 & 0.127 & 42.52 & $<$ 0.001\\

\rowcolor{lightgray} \textbf{PPR-LM (lat.)} \cite{Gauvain2004} & 8.753 & 7.265 & 6.812 & 6.534 & 27.64 & 0.016 & 0.120 & 0.107 & 0.101 & 0.097 & 24.74 & 0.007\\

\textbf{PPR-VSM} \cite{Li2007} & 8.886 & 8.990 & 9.368 & 10.056 & 46.80 & $<$ 0.001 & 0.117 & 0.120 & 0.124 & 0.133 & 37.61 & $<$ 0.001\\

\textbf{PPR-IVEC} \cite{Soufifar2011,Kesiraju2016} & 10.978 & 12.326 & 12.604 & 12.462 & 56.93 & $<$ 0.001 & 0.152 & 0.165 & 0.168 & 0.163 & 51.97 & $<$ 0.001\\
\midrule

\textbf{PPR-SM-SVM (OLR)} \cite{Shih2012} & 7.222 & 6.670 & 6.759 & 6.627 & 28.66 & 0.019 & 0.099 & 0.094 & 0.094 & 0.095 & 22.34 & 0.029\\

\rowcolor{lightgray} \textbf{PPR-SM-SVM (ODL)} & 7.222 & 6.534 & 6.534 & 6.673 & 27.64 & 0.020 & 0.099 & 0.091 & 0.093 & 0.094 & 19.78 & 0.034\\

\textbf{PPR-SM-SVM (DLM)} \cite{Lee2013} & 7.001 & 7.403 & 7.611 & 7.786 & 32.47 & 0.005 & 0.098 & 0.102 & 0.105 & 0.106 & 25.51 & 0.008\\
\midrule

\textbf{PPR-SM-SNN (OLR)} & \bf 5.974 & \bf 4.912 & 4.904 & 4.819 & 1.89 & 0.864 & \bf 0.087 & \bf 0.077 & 0.075 & 0.075 & 2.67 & 0.928\\

\rowcolor{lightgray} \textbf{PPR-SM-SNN (ODL)} & 6.349 & 5.007 & \bf 4.858 & \bf 4.728 & - & - & 0.092 & \bf 0.077 & \bf 0.074 & \bf 0.073 & - & -\\

\textbf{PPR-SM-SNN (DLM)} & 6.252 & 5.746 & 5.940 & 6.164 & 17.72 & 0.123 & 0.089 & 0.083 & 0.086 & 0.087 & 12.05 & 0.200\\
\bottomrule
\end{tabular}
\vspace{-15pt}
\end{table*}

\subsubsection{Learning strategy of PPR-SM-SNN}
\label{sec:learning_strategy_snn}
We first discuss the need to consider orthogonality when initializing the weight maps of the PPR-SM-SNN. As shown in Figs. \ref{fig:lre03_snn_curve}(a) and (b), when the initial random weight maps were set with the Glorot normal initializer without considering orthogonality \cite{Glorot2010}, the \textit{best} evaluation EER increased by approximately 8\% (from 2.157\% to 2.332\%), compared with the case involving the orthogonal initializer. Moreover, its \textit{final} (epoch 200) EER converged to 2.588\%, which is worse than 2.5\%, as shown in Fig. \ref{fig:lre03_snn_curve}(a).

Next, we discuss the drop-based learning rate schedule. As shown in Fig. \ref{fig:lre03_snn_curve}(a), without using a learning rate schedule, the curve for evaluating the EER fluctuated significantly; hence, we could not determine the stopping epoch using an additional development set or through cross-validation. Even when the learning rate was reduced from $10^{-4}$ to $10^{-5}$, as shown in Fig. \ref{fig:lre03_snn_curve}(c), the EER measure still appeared unreliable. Furthermore, we observed overfitting in Fig. \ref{fig:lre03_snn_curve}(c). However, with the drop-based learning rate schedule, when the number of epochs increases to a sufficiently large number, the learning rate will decay and barely affect the training loss and model. Hence, the EER evaluated after the last training epoch might not be the lowest, but should still be acceptable. As an example, after the 200th epoch shown in Fig. \ref{fig:lre03_snn_curve}(d), where the learning rate finally decays to $2 \times 10^{-6}$, the evaluation ERR (2.402\%) is only 11.4\% higher than the best EER (2.157\%) shown in Fig. \ref{fig:lre03_snn_curve}(a).

\begin{figure}[t]
\centering
\includegraphics[width=0.48\textwidth]{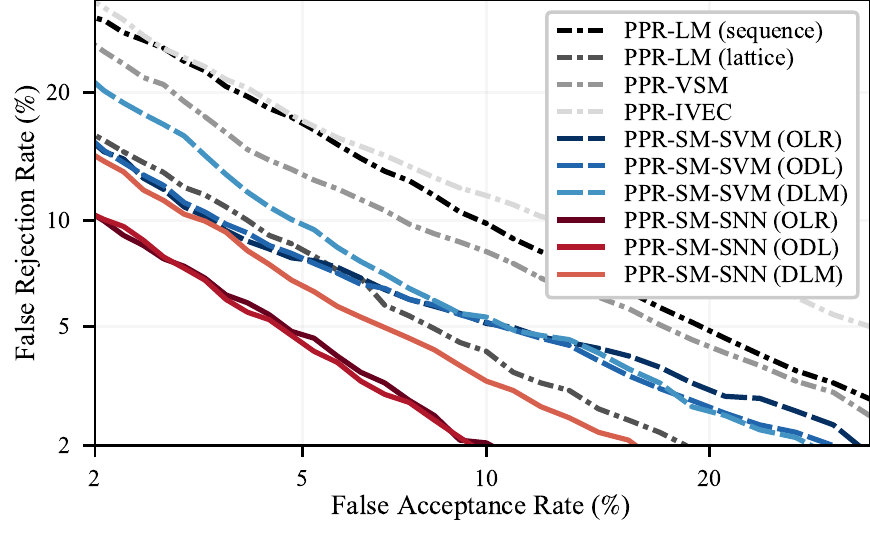}
\vspace{-10pt}
\caption{The DET curves for different methods, where each method adopts the settings achieving the best performance in Table \ref{tab:lre07_restlts}.}
\label{fig:lre07_det}
\vspace{-15pt}
\end{figure}

\subsection{Language Recognition on LRE-07}
\label{sec:experiment_lre07}
In the LRE-07 task, the model configuration, including subspace specifications and hyperparameters, such as the regularization penalty and learning rate, was predetermined by the LRE-03 task. Therefore, we used the settings for the best results in Fig. \ref{fig:lre03_oracle}. As shown in Table \ref{tab:lre07_restlts}\footnote{The baseline results may be worse than those evaluated on the same dataset reported in the corresponding papers (e.g., \cite{Ambikairajah2011}) because we have less training data available for phonotactic modeling. For example, as described in \cite{Li2013}, many systems used additional CallHome and CallFriend corpora for phonotactic modeling, but these corpora were not used in this study.} and Fig. \ref{fig:lre07_det}, lattice-based PPR-LM is superior to the other three baseline methods and comparable to PPR-SM-SVM in terms of both the EER and average cost ($C_{avg}$). Both the lattice-based PPR-LM and PPR-SM-SNN can use richer phone-contextual information to distinguish languages than the other methods. However, the performance of the PPR-SM-SNN was better than that of lattice-based PPR-LM. Regardless of the value of $n$, the EERs and $C_{avg}$ values of the PPR-SM-SNN were significantly lower than those of lattice-based PPR-LM. For example, when $n=5$, PPR-SM-SNN (ODL) achieved a 27.64\% reduction in the EER (from 6.534\% to 4.728\%) and a 24.74\% reduction in $C_{avg}$ (from 0.097 to 0.073), compared with lattice-based PPR-LM. According to Welch's t-test, the PPR-SM-SNN significantly outperformed all the other methods (with p-values $< 0.05$), as shown in Table \ref{tab:lre07_restlts}. For PPR-SM-SVM, as performed in the LRE-03 task, increasing $n$ did not improve the performance. The only exception was an increase in $n$ from 2 to 3. The results demonstrate that PPR-SM-SVM did not perform well on larger and more challenging datasets. It can be concluded that SNNs are more suitable for subspace learning than SVMs, because the \textit{reference} subspaces used in the weight maps of SNNs are not as restricted as the \textit{sample} subspaces used for kernel computation in SVMs.

\begin{figure}[t]
\centering
\includegraphics[width=0.48\textwidth]{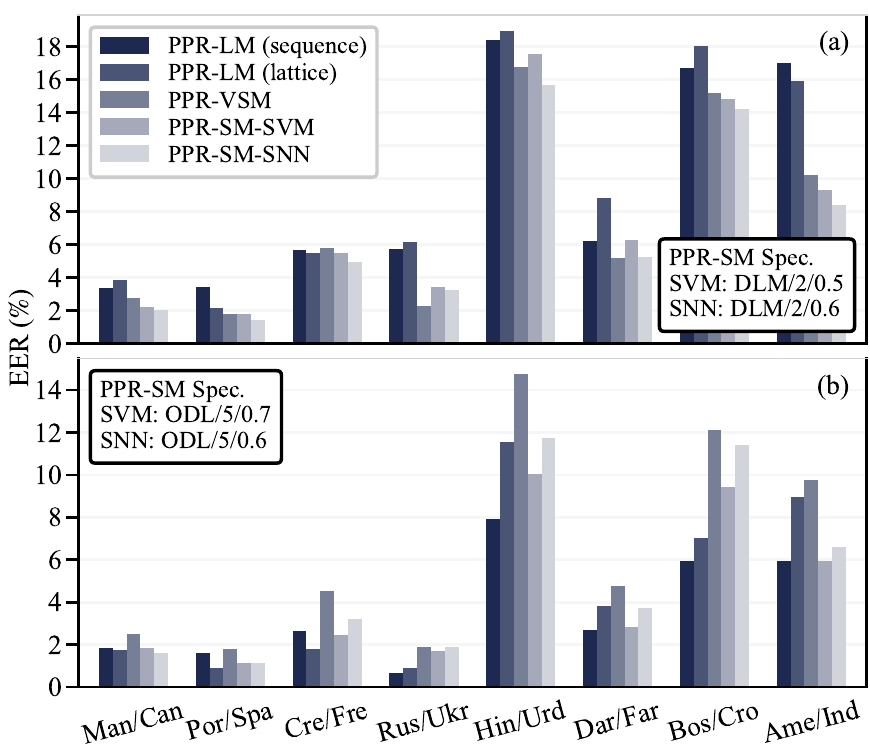}
\vspace{-10pt}
\caption{Average EERs of stratified 5-fold CV for different methods, (a) the context order $n=2$ and (b) the context order $n=5$.}
\label{fig:lre09}
\vspace{-15pt}
\end{figure}

Although LRE-07 and LRE-03 are two different tasks, we discovered that the performance trends of all the compared methods on the two tasks were generally consistent. Two new observations obtained in this study are as follows:

% 以下兩點漏看
\begin{enumerate}[wide,label=\textbf{\roman*)},labelwidth=!,labelindent=1pt]
\setlength{\itemsep}{2pt}
\item
When $n=2$, DLM generally outperformed OLR and ODL in the LRE-03 task, as shown in Fig. \ref{fig:lre03_oracle}(a). However, in the LRE-07 task, DLM did not show an advantage over OLR and ODL.
\item
Although ODL and OLR seemed to be neck and neck in most cases in Fig. \ref{fig:lre03_oracle} and Table \ref{tab:lre07_restlts}, PPR-SM-SNN (ODL) slightly outperformed PPR-SM-SNN (OLR) in Fig. \ref{fig:lre07_det}. 
\end{enumerate}

\subsection{Dialect/Accent Identification on LRE-09}
\label{sec:experiment_lre09}
In the LRE-09 task, similar to the LRE-07 task, the model configuration inherits the LRE-03 settings from the best results shown in Fig. \ref{fig:lre03_oracle}. The average EERs of the stratified five-fold cross-validation are shown in Fig. \ref{fig:lre09}. PPR-IVEC was not implemented for the task owing to its poor performance in the first two tasks. The followings were observed from Fig. \ref{fig:lre09}.

% 以下五點漏看
\begin{enumerate}[wide,label=\textbf{\roman*)},labelwidth=!,labelindent=1pt]
\setlength{\itemsep}{2pt}
\item
The last five spoken language pairs are generally considered to be mutually understandable \cite{NIST2009}, i.e., two persons who speak different languages can understand each other and communicate easily. However, the result shows that, compared with Hin/Urd, Bos/Cro, and Ame/Ind, the subtle differences in the phonotactics of Rus/Ukr and Dar/Far are easier to detect.
\item
Surprisingly, in most cases, sequence-based PPR-LM outperformed lattice-based PPR-LM, especially in the last five pairs, when longer phonotactic constraints were used (i.e., $n=5$). When $n=2$, PPR-VSM was superior to sequence-based PPR-LM and lattice-based PPR-LM, but when $n=5$, PPR-VSM was inferior. Overall, PPR-SM-SVM and PPR-SM-SNN outperformed the three baseline methods in most cases, especially when $n=2$.
\item
Contrary to the previous results, PPR-SM-SVM was better than PPR-SM-SNN when $n=5$, except for the first two pairs. One possible reason is that when the training sample set is not large enough (e.g., less than 3,000 in this task), more complex learning machines with higher capacity (such as NN-based models) will be prone to overfitting. At the representation level, more informative phonotactic tokens, such as soft-count statistics and our proposed subspaces, may not be suitable for small data volume situations.
\item
When $n=2$, PPR-VSM achieved the best results for the pairs Rus/Ukr and Dar/Far. Moreover, when $n=5$, PPR-SM-SVM outperformed PPR-SM-SNN except for the first two pairs. The results may indicate that PPR-VSM and PPR-SM-SVM, which both use SVMs as the backend classifier, have their advantages in binary classification problems. It is also obvious that the subspace-based representation is more useful than the VSM-based representation, especially when the length of the phonotactic constraint is longer.
\item
When $n=2$, regardless of the difficulty of these pairs, PPR-SM-SNN showed its superiority and stability in distinguishing dialects and accents. 
\end{enumerate}

\begin{table}[t]
\caption{EERs (\%) for LRE'03. \#PPRs denotes the number of parallel phone recognizers and the training data sets.}
\vspace{-5pt}
\label{tab:lre03}
\centering
\begin{tabular}{lcm{2cm}l}
\toprule
{\bf Methods} & {\bf EER} & \makecell[c]{\bf \#PPRs} & {\bf Training Sets} \\
\midrule
\makecell[l]{PPR-SM-SNN} & 2.157 & \makecell[c]{3 \\(BUT)} & \makecell[l]{LRE'96 \\(10 s, 30 s)} \\
\midrule
\makecell[l]{PPR-LM (lattice) \\impl. by us} & 2.269 & \makecell[c]{3 \\(BUT)} & \makecell[l]{LRE'96 \\(10 s, 30 s)} \\
\midrule
\makecell[l]{PPR-LM (lattice) \\impl. by Gauvain \cite{Gauvain2004}} & 4.000 & \makecell[c]{3 \\(Switchboard, \\CallHome)} & \makecell[l]{LRE'96 \\(3 s, 10 s, 30 s)} \\
\bottomrule
\end{tabular}
\vspace{-10pt}
\end{table}

\subsection{Comparison with Results in Literature}
Table \ref{tab:lre03} shows the results evaluated on LRE'03. We compared the results of our proposed PPR-SM-SNN and the PPR-LM (lattice) implemented by us with the result of PPR-LM (lattice) from Gauvain's paper \cite{Gauvain2004}. The main difference between our PPR-LM (lattice) and Gauvain's PPR-LM (lattice) is the use of different parallel phone recognizers. As shown in the table, our implementation of PPR-LM (lattice) is reliable; under similar conditions, its EER is lower than that of Gauvain's PPR-LM (lattice). In addition, the proposed PPR-SM-SNN method is superior to the two PPR-LM (lattice) methods. We did not find any references with reliable results of PPR-LM (lattice) performed on LRE'07 under similar training conditions. 

For PPR-VSM, we referred to Tong's work \cite{Tong2009} and Ambikairajah's work \cite{Ambikairajah2011}. The results evaluated on LRE'07 are shown in Table \ref{tab:lre07}. In \cite{Tong2009}, Tong constructed seven phone recognizers (PRs), conducted PPR-VSM based on each PR, and hence obtained EERs ranging from 8.39\% to 13.96\%. The EER of our PPR-VSM method was 8.886\%. We believe that the performance is reasonable. It is noteworthy that only one (Russian) among three languages of our PPRs (BUT) appeared in the 14 target languages in LRE'07; however, all seven languages in Tong's PRs overlapped with the 14 target languages in LRE'07. In addition, as shown in Table \ref{tab:lre07}, our PPR-VSM is worse than Ambikairajah's PPR-VSM. The reasons are twofold. First, compared with our PPR-VSM, Ambikairajah's PPR-VSM used more matched languages in PPRs. Second, the CallFriend corpus, which is a much larger training set than LRE'09 (selected), was used in Ambikairajah's work. In summary, the performance of our PPR-VSM is reasonable. It is clear that the proposed PPR-SM-SNN method is superior to our PPR-VSM baseline.

In this study, we focused on phonotactic SLR and investigated how PPRs can help distinguish languages. Therefore, we only compared the proposed methods with state-of-the-art phonotactic methods. Comparing the proposed methods with nonphonotactic systems may not be suitable for the purpose of this study. Even though the phonotactic approaches cannot compete with state-of-the-art methods based on i-vectors \cite{McCree2016} and x-vectors \cite{Snyder2018a}, they are still worthy of investigation. However, in-depth investigation and comparison of these fundamentally different methods should be performed. Moreover, because the proposed phonotactic methods and i-/x-vectors are derived based on different concepts and may contain complementary information, we will investigate unified frameworks that combine these two classes of methods in future studies.

\begin{table}[t]
\caption{EERs (\%) for LRE'07. LRE'96-05 means LRE'96, 03, and 05.}
\vspace{-5pt}
\label{tab:lre07}
\centering
\begin{tabular}{lm{1cm}m{2cm}l}
\toprule
{\bf Methods} & {\bf EER} & \makecell[c]{\bf \#PPRs} & {\bf Training Sets} \\
\midrule
\makecell[l]{PPR-SM-SNN} & 4.728 & \makecell[c]{3 \\(BUT)} & \makecell[l]{LRE'96-05, \\LRE'07 (dev), \\LRE'09 (selected)} \\
\midrule
\makecell[l]{PPR-VSM \\impl. by us} & 8.886 & \makecell[c]{3 \\(BUT)} & \makecell[l]{LRE'96-05, \\LRE'07 (dev), \\LRE'09 (selected)} \\
\midrule
\makecell[l]{PPR-VSM \\impl. by \\Tong \cite{Tong2009}} & 8.390-13.960 & \makecell[c]{7 \\(IIR-LID, \\OGI-MLTS etc.)} & \makecell[l]{LRE'96-05, \\LRE'07 (dev), \\CallFriend} \\
\midrule
\makecell[l]{PPR-VSM \\impl. by \\Ambikairajah \cite{Ambikairajah2011}} & 4.380 & \makecell[c]{7 \\(IIR-LID, \\OGI-MLTS etc.)} & \makecell[l]{LRE'07 (dev),\\ CallFriend} \\
\bottomrule
\end{tabular}
\vspace{-10pt}
\end{table}

\section{Conclusions}
\label{sec:conclusions}
Herein, we proposed a new phonotactic representation of an utterance based on the concept of linear subspace, which is neither equivalent nor reducible to a distributional or vectorial representation. It increases the capacity of the representation for containing more phonotactic information. In particular, it can utilize uncertain information produced by unreliable phone recognizers by forming phonetic vectors. In addition to the previously proposed orthogonal linear regression and dynamic linear modeling, we proposed a dictionary learning-based method for subspace construction. The resulting subspace formed under Laplace's prior condition can restore the phonotactic structure underlying an utterance; hence, it is more capable of capturing the most salient features.

In addition to SVMs, the proposed SNNs were investigated as a solution for subspace learning and experimentally proven to outperform four phonotactic baselines in most experiments, i.e., sequence-based PPR-LM, lattice-based PPR-LM, PPR-VSM, and PPR-IVEC. The input layer of the SNNs accepted subspace-shaped samples, and the intermediate weight maps were specifically designed with a specific and differentiable subspace similarity based on the principal angles. Using SNNs, the PPR-SM-SNN achieved a relative reduction of 27\% in EER on the LRE-07 language detection task, compared with the SVM-based subspace learning counterpart (PPR-SM-SVM).
	
Although we did not quantitatively analyze the time complexity, with the assistance of GPUs, the training process of SNNs was much faster than those of other lattice- or SVM-based methods compared in this study. This was particularly true when the context order exceeded 3. However, the development of SNNs can be improved further. We believe that when training SNNs, certain GPU-based realizations for setting constraints on the weight maps can stabilize the movement of the training loss and the performance of the development data. For example, the recently proposed parameterization based on the Lie group theory via an exponential map might be worth investigating \cite{Lezcano-Casado2019}.

According to the parlance of phonology, a learning approach for SLR is regarded as \textit{phonotactic} in that the smallest speech event to be addressed is a phoneme rather than an acoustic frame. Typically, the former is longer than the latter. In future studies, we will include other \textit{frame-based} statistics or information in the formation of \textit{phoneme-based} phonetic vectors, such as the PLLR \cite{Diez2012,DHaro2014} and frame-based features  \cite{Han2012}. The integration of the proposed subspace approaches with the PLLR features will be an interesting future direction.

\appendix

\begin{algorithm}[H]
\caption{The Subspace Method for DLMs}
\label{alg:dlm}
\begin{algorithmic}[1]
\REQUIRE The observation matrix $\mathbf{Y}=[\mathbf{y}_1,\dotsc,\mathbf{y}_K]$.
\ENSURE The estimates $\hat{\mathbf{A}}$ and $\hat{\mathbf{C}}$.
\STATE Perform truncated SVD on $\mathbf{Y}$, such that $\mathbf{Y}\approx\mathbf{U}\mathbf{\Sigma}\mathbf{V}^T$ \cite{Chu2001}. $\mathbf{\Sigma}$ is a $d\times d$ diagonal matrix containing the largest $d$ positive singular values. $\mathbf{U}\in\mathbb{R}^{M\times d}$ and $\mathbf{V}\in\mathbb{R}^{K\times d}$ are the corresponding eigen-matrices.
\STATE Set $\hat{\mathbf{C}}$, the estimate of $\mathbf{C}$, to be $\mathbf{U}$.
\STATE Set $\hat{\mathbf{X}}_{1:K}$, the estimate of the state matrix $\mathbf{X}_{1:K}$, to be $\mathbf{\Sigma} \mathbf{V}^T$, where $\hat{\mathbf{X}}_{1:K}$ stands for $[\hat{\mathbf{x}}_1,\dotsc,\hat{\mathbf{x}}_K]\in\mathbb{R}^{d\times K}$.
\STATE Estimate $\hat{\mathbf{A}}$ by solving the following equation:
\begin{subequations}
\label{eq:olr_dlm}
\begin{align}
\hat{\mathbf{A}}=\argmin_{\mathbf{A}}\sum_{k=1}^{K-1}\|\hat{\mathbf{x}}_{k+1}-\mathbf{A}\hat{\mathbf{x}}_k\|^2\\
\text{subject to}~\mathbf{A}^T\mathbf{A}=\mathbf{I}_d,
\end{align}
\end{subequations}
where $\|\cdot\|$ denotes the $\mathit{l}^2$-norm.
\end{algorithmic}
\end{algorithm}

Algorithm \ref{alg:dlm} differs from the algorithms in our previous study \cite{Lee2013} and other related studies \cite{Ljung1999,Hamm2008} in that an orthonormal constraint on $\mathbf{A}$ is added in the last step. Such a constraint is necessary for the pursuit of orthogonality presented in Section \ref{sec:orthogonality_pursuit}. In fact, (\ref{eq:olr_dlm}) is an orthogonal procrustes problem, and its solution is an orthonormal matrix $\hat{\mathbf{A}}$ that most closely maps $\hat{\mathbf{X}}_{1:(K-1)}$ to $\hat{\mathbf{X}}_{2:K}$. By referring to \cite[Section~1]{Chu2001a}, the process of solving (\ref{eq:olr_dlm}) can be summarized as Algorithm \ref{alg:opp}.

\begin{algorithm}[H]
\caption{The Orthogonal Procrustes Problem}
\label{alg:opp}
\begin{algorithmic}[1]
\REQUIRE The state matrix $\hat{\mathbf{X}}_{1:K}$ estimated by Algorithm \ref{alg:dlm}.
\ENSURE The estimate $\hat{\mathbf{A}}$.
\STATE Perform full SVD on $\hat{\mathbf{X}}_{2:K}\hat{\mathbf{X}}^T_{1:(K-1)}$, such that $\hat{\mathbf{X}}_{2:K}\hat{\mathbf{X}}^T_{1:(K-1)}=\mathbf{U}\mathbf{\Sigma}\mathbf{V}^T$, where $\mathbf{U}$, $\mathbf{\Sigma}$, and $\mathbf{V}$ are all $d\times d$ matrices.
\STATE Set $\hat{\mathbf{A}}$, the estimate of $\mathbf{A}$, to be $\mathbf{U}\mathbf{V}^T$.
\end{algorithmic}
\end{algorithm}

To solve (\ref{eq:odl}) iteratively, two propositions must be provided:

\begin{prop}
\label{prop:1}
Given $\mathbf{S}^T\mathbf{S}=\mathbf{I}_d$, $\hat{\mathbf{W}}=O_{\lambda_{odl}}(\mathbf{S}^T\mathbf{Z})$ is the unique solution of the following minimization problem:
\begin{equation}
\min\limits_{\mathbf{W}}\|\mathbf{Z}-\mathbf{S}\mathbf{W}\|^2_F+\lambda_{odl}^2\|\mathbf{W}\|_0.
\end{equation}
$O_{\lambda_{odl}}(\cdot)$ is a threshold operator. $[O_{\lambda_{odl}}(\mathbf{X})]_{ij}=\mathbf{X}_{ij}$ if the absolute value of the $(i, j)$-th entry of $\mathbf{X}$ is larger than $\lambda$, and $[O_{\lambda_{odl}}(\mathbf{X})]_{ij}=0$ otherwise.
\end{prop}

\begin{prop}
\label{prop:2}
$\hat{\mathbf{S}}_{odl}=\mathbf{P}\mathbf{Q}^T$ is the unique solution of the following minimization problem:
\begin{subequations}
\begin{align}
\min\limits_{\mathbf{S}}\|\mathbf{Z}-\mathbf{S}\hat{\mathbf{W}}\|^2_F\\
\text{subject to}~\mathbf{S}^T\mathbf{S}=\mathbf{I}_d.
\end{align}
\end{subequations}
$\mathbf{P}$ and $\mathbf{Q}$ denote the orthonormal matrices defined in the following truncated SVD of $\mathbf{Z}\hat{\mathbf{W}}^T$:
\begin{equation}
\mathbf{Z}\hat{\mathbf{W}}^T=\mathbf{P}\mathbf{\Sigma}\mathbf{Q}^T.
\end{equation}
\end{prop}

The mathematical proof is provided in \cite{Bao2013}. We performed some minor modifications to suit to our study, but the proof is still valid. The process of solving (\ref{eq:odl}) is provided in Algorithm \ref{alg:solr}.

\begin{algorithm}[H]
\caption{The Orthogonal Dictionary Learning}
\label{alg:solr}
\begin{algorithmic}[1]
\REQUIRE
The contextualized phonetic matrix $\mathbf{Z}\in\mathbb{R}^{D\times K}$.
\ENSURE
The subspace $\hat{\mathbf{S}}_{odl}\in\mathbb{R}^{D\times d}$.
\STATE
Set a feasible initial guess $\mathbf{S}^{(0)}$. 
\STATE
For $j=0,1,...,J$, \\
\hskip\algorithmicindent (a) $\mathbf{W}^{(j)}\gets O_{\lambda_{odl}}(\mathbf{S}^T\mathbf{Z})$; \\
\hskip\algorithmicindent (b) Perform truncated SVD on $\mathbf{Z}\mathbf{W}^{(j)T}=\mathbf{P}\mathbf{\Sigma}\mathbf{Q}^T$; \\
\hskip\algorithmicindent (c) $\mathbf{S}^{(i+1)}\gets \mathbf{P}\mathbf{Q}^T$.
\STATE
$\hat{\mathbf{S}}_{odl}\gets \mathbf{S}^{(J+1)}$.
\end{algorithmic}
\end{algorithm}

In this study, $\mathbf{S}^{(0)}$ was horizontally stacked by $d$ randomly sampled column vectors of an identity matrix $\mathbf{I}_D\in\mathbb{R}^{D\times D}$.

\ifCLASSOPTIONcaptionsoff
  \newpage
\fi

\bibliographystyle{IEEEtran}
\bibliography{references.bib}

\end{document}